\documentclass[twoside,11pt]{article}

\usepackage{submodules/symbols/ps}
\hypersetup{
    pdftitle={Graph Coloring to Reduce Computation Time in Prioritized Planning},
    pdfauthor={Scheffe, Kahle, and Alrifaee},
    pdfpagelayout=SinglePage,
}
\glsdisablehyper %
\acsetup{make-links=false}
\newcommand{\anAgent}{\ensuremath{i}}
\newcommand{\anotherAgent}{\ensuremath{j}}
\newcommand{\timestep}{\ensuremath{k}}
\newcommand{\timestepIterator}{\ensuremath{l}}

\newcommand{\ofAgent}[1]{\ensuremath{^{(#1)}}}
\newcommand{\ofAgentInAgent}[2]{\ensuremath{^{(#1 \,\vert\, #2)}}}
\newcommand{\forTimeAtTime}[2]{\ensuremath{_{#1 \,\vert\, #2}}}

\newcommand{\seqCoupling}{\ensuremath{\leftarrow}}

\newglossaryentry{matrix:Adjacency}{
	name=\ensuremath{\bm{D}},
	description={Adjacency matrix},
	sort={D},
    type=symbol
}
\newcommand{\matAdjacency}{\gls{matrix:Adjacency}}
\newcommand{\matAdjacencyElement}[1]{\glslink{matrix:Adjacency}{\matAdjacency_{#1}}}

\newglossaryentry{set:realNumbers}{
	name=\ensuremath{\mathbb{R}},
	description={Set of real numbers},
	sort={real numbers},
    type=symbol
}
\newcommand{\setRealNumbers}{\gls{set:realNumbers}}

\newglossaryentry{set:naturalNumbers}{
	name=\ensuremath{\mathbb{N}},
	description={Set of natural numbers},
	sort={natural numbers},
    type=symbol
}
\newcommand{\setNaturalNumbers}{\gls{set:naturalNumbers}}

\newglossaryentry{set:systemStates}{
	name=\ensuremath{\mathcal{S}},
	description={Set of system states},
	sort={set of system states},
    type=symbol
}

\newglossaryentry{set:bigO}{
	name=\ensuremath{O},
	description={Big O},
	sort={O},
    type=symbol
}

\newglossaryentry{scalar:Weight}{
	name=\ensuremath{w},
	description={Weight},
	sort={weight},
    type=symbol
}

\newglossaryentry{scalar:NumberOfAgents}{
    name=\ensuremath{N_A},
    description={Number of agents},
    sort={Number of agents},
    type=symbol
}
\newcommand{\numAgents}{\gls{scalar:NumberOfAgents}}

\newglossaryentry{graph:path}{
    name=\ensuremath{\pi},
    description={Path},
    sort={Path},
    type=symbol
}
\newcommand{\graphPath}{\gls{graph:path}}

\newglossaryentry{scalar:NumberOfVerticesInPath}{
    name=\ensuremath{N_{\graphPath}},
    description={Number of vertices in path $\graphPath$, or length},
    sort={Number of vertices in path},
    type=symbol
}

\newglossaryentry{sym:horizonControl}{
	name=\ensuremath{N_u},
	description={Control horizon in model predictive control},
	sort={Nu},
    type=symbol
}

\newglossaryentry{sym:horizonPrediction}{
	name=\ensuremath{N_p},
	description={Prediction horizon in model predictive control},
	sort={Np},
    type=symbol
}
\newcommand{\horizonPrediction}{\gls{sym:horizonPrediction}}

\newglossaryentry{sym:vehicleOrientation}{
	name=\ensuremath{\psi},
	description={Vehicle orientation},
	sort={psi},
    type=symbol
}

\newglossaryentry{sym:sysModelContinuous}{
    name=\ensuremath{f},
    description={Continuous-time system model},
    sort={f continuous-time},
    type=symbol
}
\newcommand{\sysModelContinuous}{\gls{sym:sysModelContinuous}}

\newglossaryentry{sym:sysModelDiscrete}{
    name=\ensuremath{f_{d}},
    description={Discrete-time system model},
    sort={f discrete-time},
    type=symbol
}
\newcommand{\sysModelDiscrete}{\gls{sym:sysModelDiscrete}}

\newglossaryentry{sym:sysControlInputs}{
	name=\ensuremath{\bm{u}},
	description={System control inputs},
	sort=u,
    type=symbol
}
\NewDocumentCommand{\sysControlInputs}{ o }{\glslink{sym:sysControlInputs}{%
    \IfNoValueTF{#1}%
        {\ensuremath{\bm{u}}}%
        {\ensuremath{\bm{u}\ofAgent{#1}}}%
}}

\newglossaryentry{sym:outputs}{
	name=\ensuremath{\bm{y}},
	description={System outputs},
	sort={y},
    type=symbol
}

\newglossaryentry{sym:sysSpeed}{
	name=\ensuremath{\mathrm{v}},
	description={Vehicle speed},
	sort={v},
    type=symbol
}
\newcommand{\sysSpeed}{\gls{sym:sysSpeed}}

\newglossaryentry{sym:inSpeed}{
	name=\ensuremath{u_{\sysSpeed}},
	description={Vehicle input speed},
	sort={uv},
    type=symbol
}

\newglossaryentry{sym:steering-angle}{
	name=\ensuremath{\delta},
	description={Vehicle steering angle},
	sort={delta},
    type=symbol
}

\newglossaryentry{sym:inSteering}{
	name=\ensuremath{u_{\delta}},
	description={Vehicle input steering angle},
	sort={ud},
    type=symbol
}

\newglossaryentry{sym:nColors}{
	name=\ensuremath{N_c},
	description={Number of colors},
	sort={Number of colors},
    type=symbol
}
\newcommand{\numColors}{\gls{sym:nColors}}

\newglossaryentry{sym:nStates}{
	name=\ensuremath{n},
	description={Number of states of a dynamical system},
	sort={Number of states},
    type=symbol
}
\newcommand{\numStates}{\gls{sym:nStates}}

\newglossaryentry{sym:nInputs}{
    name=\ensuremath{m},
    description={Number of inputs of a dynamical system},
    sort={m number of inputs},
    type=symbol
}
\newcommand{\numInputs}{\gls{sym:nInputs}}

\newglossaryentry{sym:nLevels}{
	name=\ensuremath{N_{c}},
	description={Number of computation levels},
	sort={Number of computation levels},
    type=symbol
}
\newcommand{\numLevels}{\gls{sym:nLevels}}

\newglossaryentry{sym:nLevelsAllowed}{
	name=\ensuremath{N_{\text{CL,al.}}},
	description={Allowed number of computation levels},
	sort={Number of computation levels allowed},
    type=symbol
}

\newglossaryentry{sym:numGroups}{
	name=\ensuremath{N_{g}},
	description={Number of parallelly computing groups of agents},
	sort={Number of groups},
    type=symbol
}

\newglossaryentry{sym:fcnPrio}{
    name=\ensuremath{p},
    description={Prioritization function},
    sort={Prioritization function},
    type=symbol
}
\newcommand{\fcnPrio}{\gls{sym:fcnPrio}}

\newglossaryentry{sym:tSample}{
	name=\ensuremath{T_s},
	description={Sample Time},
	sort={T sample},
    type=symbol
}

\newglossaryentry{sym:tSolve}{
	name=\ensuremath{T_\text{sol.}},
	description={Computation time \tSolveB{\anAgent} that agent $\anAgent$ needs to solve its \ac{ocp}},
	sort={T solve},
    type=symbol
}

\newcommand{\tSolveB}[1]{\glslink{sym:tSolve}{\ensuremath{\ensuremath{T_\text{sol.}}^{(#1)}}}}

\newglossaryentry{sym:tSolveUpper}{
	name=\ensuremath{T_\text{sol.,max}},
	description={Upper computation time $T_\text{sol.,max}\ofAgent{\anAgent}$ that agent $\anAgent$ needs to solve it \ac{ocp}},
	sort={T solve upper},
    type=symbol
}

\newglossaryentry{sym:vertices}{
	name=\ensuremath{\mathcal{V}},
	description={Set of vertices},
	sort={Vertices},
    type=symbol
}
\newcommand{\setVertices}{\gls{sym:vertices}}
\newcommand{\setAgents}{\setVertices}

\newcommand{\helpSetPredecessors}[1]{\ensuremath{\setVertices^{(#1\leftarrow)}}}
\newglossaryentry{sym:predecessors}{
	name=\ensuremath{\helpSetPredecessors{i}},
	description={Set of predecessors of vertex $i$},
	sort={Vertices 1},
    type=symbol
}
\newcommand{\setPredecessors}[1]{\glslink{sym:predecessors}{\ensuremath{\helpSetPredecessors{#1}}}}

\newcommand{\helpSetPredecessorsPar}[1]{\ensuremath{\setVertices^{(#1\leftarrow)}_{\text{par.}}}}
\newglossaryentry{sym:predecessorsPar}{
	name=\ensuremath{\helpSetPredecessorsPar{i}},
	description={Set of predecessors of vertex $i$ that have parallel couplings with it},
	sort={Vertices 2},
    type=symbol
}

\newcommand{\helpSetPredecessorsSeq}[1]{\ensuremath{\setVertices^{(#1\leftarrow)}_{\text{seq.}}}}
\newglossaryentry{sym:predecessorsSeq}{
	name=\ensuremath{\helpSetPredecessorsSeq{i}},
	description={Set of predecessors of vertex $i$ that have sequential couplings with it},
	sort={Vertices 3},
    type=symbol
}

\newcommand{\helpSetSuccessors}[1]{\ensuremath{\setVertices^{(#1\rightarrow)}}}
\newglossaryentry{sym:successors}{
	name=\ensuremath{\helpSetSuccessors{i}},
	description={Set of successors of vertex $i$},
	sort={Vertices 4},
    type=symbol
}
\newcommand{\setSuccessors}[1]{\glslink{sym:successors}{\ensuremath{\helpSetSuccessors{#1}}}}

\newglossaryentry{sym:neighbors}{
	name=\ensuremath{\setVertices^{(i)}},
	description={Set of neighbors of vertex $i$},
	sort={Vertices 0},
    type=symbol
}
\newcommand{\setNeighbors}[1]{\glslink{sym:neighbors}{\ensuremath{\setVertices^{(#1)}}}}

\newglossaryentry{sym:degree}{
	name=\ensuremath{d^{(i)}},
	description={Degree of vertex $i$. Sum of in-degree and out-degree},
	sort=degree,
    type=symbol
}
\newcommand{\vertexDegree}[1]{\glslink{sym:degree}{\ensuremath{d^{(#1)}}}}

\newcommand{\helpVertexInDegree}[1]{\ensuremath{d^{(#1\leftarrow)}}}
\newglossaryentry{sym:inDegree}{
    name=\helpVertexInDegree{i},
    description={In-degree of vertex $i$},
    sort={degree in},
    type=symbol
}
\newcommand{\vertexInDegree}[1]{\glslink{sym:inDegree}{\helpVertexInDegree{#1}}}

\newcommand{\helpVertexOutDegree}[1]{\ensuremath{d^{(#1\rightarrow)}}}
\newglossaryentry{sym:outDegree}{
    name=\helpVertexOutDegree{i},
    description={Out-degree of vertex $i$},
    sort={degree out},
    type=symbol,
}
\newcommand{\vertexOutDegree}[1]{\glslink{sym:outDegree}{\helpVertexOutDegree{#1}}}

\newglossaryentry{sym:matLevels}{
	name=\ensuremath{\bm{L}},
	description={Matrix of computation levels},
	sort=L,
    type=symbol
}

\newglossaryentry{sym:tComp}{
	name=\ensuremath{T},
	description={Computation time},
	sort={T},
    type=symbol
}
\newcommand{\tComp}{\gls{sym:tComp}}

\newglossaryentry{sym:tCompNcs}{
	name=\ensuremath{T},
	description={Computation time of \iac{ncs}},
	sort={T},
    type=symbol
}
\newcommand{\tCompNcs}{\gls{sym:tCompNcs}}

\newglossaryentry{graph:Undirected}{
	name=\ensuremath{\mathcal{G}},
	description={Undirected Graph},
	sort={graph1},
    type=symbol
}
\newcommand{\graphUndirected}{\gls{graph:Undirected}}

\newglossaryentry{graph:Directed}{
    name=\ensuremath{\vec{\gls*{graph:Undirected}}},
	description={Directed Graph},
	sort={graph2},
    type=symbol
}
\newcommand{\graphDirected}{\gls{graph:Directed}}

\newglossaryentry{mat:edgeUtilities}{
    name=\ensuremath{M_\text{u}},
	description={Edge utility matrix},
	sort={matrix edge utilities},
    type=symbol
}

\newglossaryentry{sym:setColors}{
    name=\ensuremath{\mathcal{C}},
    description={Set of colors},
    sort=Colors,
    type=symbol
}
\newcommand{\setColors}{\gls{sym:setColors}}

\newglossaryentry{sym:varControlInvariantSet}{
	name=\ensuremath{\mathcal{C}_{\text{inv}}},
	description={Control invariant set},
	sort={Control invariant set},
    type=symbol
}

\newglossaryentry{set:Weights}{
	name=\ensuremath{\mathcal{W}},
	description={Set of weights in a weighted graph},
    sort={Weights},
    type=symbol
}

\newglossaryentry{set:Edges}{
	name=\ensuremath{\mathcal{E}},
	description={Set of edges; used to indicate that only undirected edges exist},
    sort={Edges},
    type=symbol
}
\newcommand{\setEdges}{\gls{set:Edges}}

\newglossaryentry{sym:varEdgeUndirected}{
	name=\ensuremath{(i - j)},
	description={Undirected between vertex $i$ and vertex $j$},
	sort={edge1},
    type=symbol
}
\newcommand{\edgeUndirected}[2]{\glslink{sym:varEdgeUndirected}{\ensuremath{(#1 - #2)}}}

\newglossaryentry{sym:setEdgesDirected}{
	name=\ensuremath{\vec{\gls*{set:Edges}}},
	description={Set of directed edges},
	sort={Edges directed},
    type=symbol
}
\newcommand{\setEdgesDirected}{\gls{sym:setEdgesDirected}}

\newglossaryentry{sym:varEdge}{
	name=\ensuremath{(i \rightarrow j)},
	description={Directed edge from vertex $i$ to vertex $j$},
	sort={edge2},
    type=symbol
}
\newcommand{\edgeDirected}[2]{\glslink{sym:varEdge}{\ensuremath{(#1 \rightarrow #2)}}}

\newglossaryentry{sym:fnReorder}{
	name=\ensuremath{f_r},
	description={Reordering function for graph color values},
	sort=fr,
    type=symbol
}

\newglossaryentry{sym:fcnObjective}{
    name=\ensuremath{J},
    description={Objective function of an optimization problem},
    sort=J,
    type=symbol
}
\NewDocumentCommand{\fcnObjective}{ o }{\glslink{sym:fcnObjective}{%
    \IfNoValueTF{#1}%
        {\ensuremath{J}}%
        {\ensuremath{J^{(#1)}}}%
}}
\NewDocumentCommand{\fcnObjectiveNcs}{ o }{\glslink{sym:fcnObjective}{%
    \IfNoValueTF{#1}%
    {\ensuremath{J}}%
        { \ensuremath{J_{#1}} }%
}}

\newglossaryentry{sym:fcnObjectiveState}{
    name=\ensuremath{\ell_{x}},
    description={Reference tracking objective function},
    sort={lx Reference tracking objective function},
    type=symbol
}
\NewDocumentCommand{\fcnObjectiveState}{ o }{\glslink{sym:fcnObjectiveState}{%
    \IfNoValueTF{#1}%
        {\ensuremath{\ell_{x}}}%
        {\ensuremath{\ell_{x}^{(#1)}}}%
}}

\newglossaryentry{sym:fcnObjectiveStateTerminal}{
    name=\ensuremath{\ell_{x,f}},
    description={Reference tracking objective terminal function},
    sort={lf Reference tracking objective terminal function},
    type=symbol
}
\NewDocumentCommand{\fcnObjectiveStateTerminal}{ o }{\glslink{sym:fcnObjectiveStateTerminal}{%
    \IfNoValueTF{#1}%
        {\ensuremath{\ell_{x,f}}}%
        {\ensuremath{\ell_{x,f}^{(#1)}}}%
}}

\newglossaryentry{sym:fcnObjectiveInput}{
    name=\ensuremath{\ell_{u}},
    description={Input change objective function},
    sort={lu Input change objective function},
    type=symbol
}
\NewDocumentCommand{\fcnObjectiveInput}{ o }{\glslink{sym:fcnObjectiveInput}{%
    \IfNoValueTF{#1}%
        {\ensuremath{\ell_{u}}}%
        {\ensuremath{\ell_{u}^{(#1)}}}%
}}

\newglossaryentry{sym:fcnObjectiveCoupling}{
    name=\ensuremath{\ell_\text{c}},
    description={Coupling objective function},
    sort={lc Coupling objective function},
    type=symbol
}
\NewDocumentCommand{\fcnObjectiveCoupling}{ oo }{\glslink{sym:fcnObjectiveCoupling}{%
    \IfNoValueTF{#1}%
        {\ensuremath{\ell_\text{c}}}%
        {\ensuremath{\ell_\text{c}^{(#1,#2)}}}%
}}

\newglossaryentry{sym:fcnConstraintCoupling}{
    name=\ensuremath{c_\text{c}},
    description={Coupling constraint function},
    sort={cc Coupling constraint function},
    type=symbol
}
\NewDocumentCommand{\fcnConstraintCoupling}{ oo }{\glslink{sym:fcnConstraintCoupling}{%
    \IfNoValueTF{#1}%
        {\ensuremath{c_\text{c}}}%
        {\ensuremath{c_\text{c}^{(#1,#2)}}}%
}}

\newglossaryentry{sym:prediction}{
	name=\ensuremath{\bm{x}\ofAgentInAgent{j}{i}\forTimeAtTime{\cdot}{k}},
	description={Prediction of agent $j$ in agent $i$ at time $k$},
	sort=x,
    type=symbol,
}
\NewDocumentCommand{\agentPrediction}{ oo }{\glslink{sym:prediction}{%
    \IfNoValueTF{#1}%
        {\ensuremath{ \bm{x}\forTimeAtTime{\cdot}{\timestep} }}%
        {\IfNoValueTF{#2}%
            { \ensuremath{ \bm{x}\forTimeAtTime{\cdot}{\timestep}\ofAgent{#1} } }%
            { \ensuremath{ \bm{x}\forTimeAtTime{\cdot}{\timestep}\ofAgentInAgent{#1}{#2} } }%
        }
}}

\newglossaryentry{sym:state}{
	name=\ensuremath{\bm{x}},
	description={System state},
	sort=x,
    type=symbol
}
\newcommand{\sysState}{\gls{sym:state}}

\newglossaryentry{sym:stateAgent}{
	name=\ensuremath{\sysState^{(i)}_{(k)}},
	description={System state of agent $i$ at time $k$},
	sort=x,
    type=symbol,
}

\newcommand{\sysStateAgentTime}[2]{\glslink{sym:stateAgent}{\ensuremath{\sysState^{(#1)}_{#2}}}}

\newglossaryentry{sym:ref}{
	name=\ensuremath{\bm{r}^{(i)}_{k}},
	description={System state reference of agent $i$ at time $k$},
	sort=x ref,
    type=symbol
}

\newglossaryentry{sym:setReachable}{
	name=\ensuremath{\mathcal{R}^{(i)}},
	description={reachable set of agent $i$},
	sort={Reachable set},
    type=symbol
}
\newcommand{\setReachable}{\glslink{sym:setReachable}{\ensuremath{\mathcal{R}}}}

\newglossaryentry{set:occupiedArea}{
	name=\ensuremath{\mathcal{O}^{(i)}},
	description={Set of the occupied area of the predicted trajectory of agent $\anAgent$},
	sort={occupied area},
    type=symbol
}

\newglossaryentry{set:feasibleStates}{
	name=\ensuremath{\mathcal{X}},
	description={set of feasible states},
	sort={x},
    type=symbol
}
\newcommand{\setFeasibleStates}{\gls{set:feasibleStates}}

\newglossaryentry{set:feasibleStatesTerminal}{
	name=\ensuremath{\mathcal{X}_f},
	description={set of feasible states at the prediction horizon},
	sort={x},
    type=symbol
}

\newglossaryentry{set:feasibleInputs}{
	name=\ensuremath{\mathcal{U}},
	description={set of feasible inputs},
	sort={u},
    type=symbol
}
\newcommand{\setFeasibleInputs}{\gls{set:feasibleInputs}}

\newglossaryentry{sym:numStatesConfSpace}{
    name=\ensuremath{n_p},
    description={Number of states that are in the conflictual space},
    sort={n number of states that are in the conflictual space},
    type=symbol
}

\newglossaryentry{sym:fnProj}{
    name=\text{proj},
    description={A function that projects a reachable set of system states in the conflictual space},
    sort={Project function},
    type=symbol
}

\newglossaryentry{sym:setClasses}{
	name=\ensuremath{\mathfrak{V}},
	description={Set of classes of parallelizable agents},
	sort={set class agents},
    type=symbol
}

\newglossaryentry{sym:classAgents}{
	name=\ensuremath{\setVertices_{c}},
	description={Class of parallelizable agents},
	sort={class agents},
    type=symbol
}
\NewDocumentCommand{\classAgents}{ o }{\glslink{sym:classAgents}{%
    \IfNoValueTF{#1}%
        {\ensuremath{\setVertices_{c}}}%
        {\ensuremath{\setVertices_{#1}}}%
}}

\newglossaryentry{sym:sequenceClasses}{
	name=\ensuremath{(s_z)_{z=1}^{\numLevels}},
	description={Sequence of classes of parallelizable agents},
	sort={sequence class agents},
    type=symbol
}

\newglossaryentry{sym:matSchedule}{
	name=\ensuremath{\bm{S}},
	description={Matrix of computation schedule},
	sort=S,
    type=symbol
}

\newglossaryentry{sym:setSequences}{
	name=\ensuremath{\mathcal{S}},
	description={Set of sequences},
	sort={Set Sequences},
    type=symbol
}

\newglossaryentry{sym:tSolveClass}{
	name=\ensuremath{T_\text{sol.}},
	description={Computation time \tSolveClass{\classAgents} that a agent class $\classAgents$ needs to solve their \acp{ocp}},
	sort={T solve class},
    type=symbol
}
\newcommand{\tSolveClass}[1]{\glslink{sym:tSolveClass}{\ensuremath{\ensuremath{T_\text{sol.}}^{[#1]}}}}

\newglossaryentry{sym:tCompNcsUpper}{
	name=\ensuremath{\hat{T}_{\text{NCS}}},
	description={Fixed computation time for \iac{ncs}},
	sort={T NCS fixed},
    type=symbol
}

\newglossaryentry{sym:latinSquare}{
	name=\ensuremath{\mathfrak{L}},
	description={Latin Square},
	sort={Latin Square},
    type=symbol
}

\DeclareAcronym{admm}{
    short = ADMM,
    long  = Alternating Direction Method of Multipliers
}
\DeclareAcronym{cav}{
    short = CAV,
    long  = connected and automated vehicle,
}
\DeclareAcronym{cbs}{
    short = CBS,
    long  = conflict-based search,
}
\DeclareAcronym{cg}{
    short = CG,
    long  = center of gravity
}
\DeclareAcronym{cdmpc}{
    short = coop. DMPC,
    long  = cooperative distributed \acl{mpc}
}
\DeclareAcronym{cmpc}{
    short = CMPC,
    long  = centralized \acl{mpc}
}
\DeclareAcronym{cpm}{
    short = CPM,
    long  = Cyber-Physical Mobility
}
\DeclareAcronym{cpmlab}{
    short = CPM Lab,
    long  = \acl{cpm} Lab
}
\DeclareAcronym{ctg}{
    short = CTG,
    long  = cost to go
}
\DeclareAcronym{ctc}{
    short = CTC,
    long  = cost to come
}
\DeclareAcronym{dag}{
    short = DAG,
    long  = directed acyclic graph
}
\DeclareAcronym{dds}{
    short = DDS,
    long  = data distribution service
}
\DeclareAcronym{dmpc}{
    short = DMPC,
    long  = distributed \acl{mpc}
}
\DeclareAcronym{dpc}{
    short = DPC,
    long  = distributed predictive control
}
\DeclareAcronym{fsa}{
    short = FSA,
    long  = finite state automaton,
    short-indefinite = an,
}
\DeclareAcronym{fca}{
    short = FCA,
    long = future collision assessment,
    short-indefinite = an,
}
\DeclareAcronym{ffo}{
    short = FFO,
    long  = first-fit ordering,
    short-indefinite = an,
}
\DeclareAcronym{fov}{
    short = FOV,
    long  = field of view,
    short-indefinite = an,
}
\DeclareAcronym{fpv}{
    short = FPV,
    long  = first-person view,
    short-indefinite = an,
}
\DeclareAcronym{hdv}{
    short = HDV,
    long  = human-driven vehicle,
    short-indefinite = an,
}
\DeclareAcronym{hil}{
    short = HiL,
    long  = hardware-in-the-loop,
}
\DeclareAcronym{hlc}{
    short = HLC,
    long  = high-level controller,
    short-indefinite = an,
}
\DeclareAcronym{ldo}{
    short = LDO,
    long  = largest degree ordering,
    short-indefinite = an,
}
\DeclareAcronym{llc}{
    short = LLC,
    long  = low-level controller,
    short-indefinite = an,
}
\DeclareAcronym{lwa}{
    short = LWA*,
    long  = lazy weighted A*,
    short-indefinite = an,
}
\DeclareAcronym{lsp}{
    short = LazySP,
    long  = lazy shortest path,
    short-indefinite = an,
}
\DeclareAcronym{lra}{
    short = LRA*,
    long  = lazy receding horizon A*,
    short-indefinite = an,
}
\DeclareAcronym{mamp}{
    short = MAMP,
    long  = multi-agent motion planning,
    short-indefinite = a,
}
\DeclareAcronym{mapf}{
    short = MAPF,
    long  = multi-agent path finding,
    short-indefinite = a,
}
\DeclareAcronym{mas}{
    short = MAS,
    long  = multi-agent system,
    short-indefinite = an,
}
\DeclareAcronym{mcts}{
    short = MCTS,
    long  = Monte Carlo tree search,
    short-indefinite = an,
}
\DeclareAcronym{mip}{
    short = MIP,
    long  = mixed integer programming,
    short-indefinite = an,
}
\DeclareAcronym{mil}{
    short = MiL,
    long  = model-in-the-loop,
}
\DeclareAcronym{milp}{
    short = MILP,
    long  = mixed integer linear programming,
    short-indefinite = an,
}
\DeclareAcronym{mld}{
    short = MLD,
    long  = mixed logical dynamical,
    short-indefinite = an,
}
\DeclareAcronym{mlc}{
    short = MLC,
    long  = mid-level controller,
    short-indefinite = an,
}
\DeclareAcronym{mp}{
    short = MP,
    long  = motion primitive,
    short-indefinite = an,
}
\DeclareAcronym{mpa}{
    short = MPA,
    long  = motion primitive automaton,
    short-indefinite = an,
}
\DeclareAcronym{mpc}{
    short = MPC,
    long  = model predictive control,
    short-indefinite = an,
}
\DeclareAcronym{ncs}{
    short = NCS,
    long  = networked control system,
    short-indefinite = an,
}
\DeclareAcronym{nlp}{
    short = NLP,
    long  = nonlinear programming,
    short-indefinite = an,
}
\DeclareAcronym{ocp}{
    short = OCP,
    long  = optimal control problem,
    short-indefinite = an,
    long-indefinite = an,
}
\DeclareAcronym{odd}{
    short = ODD,
    long  = operational design domain,
    short-indefinite = an,
    long-indefinite = an,
}
\DeclareAcronym{ode}{
    short = ODE,
    long  = ordinary differential equation,
    short-indefinite = an,
    long-indefinite = an,
}
\DeclareAcronym{pdmpc}{
    short = \mbox{P-DMPC},
    long  = prioritized \acl{dmpc}
}
\DeclareAcronym{pil}{
    short = PiL,
    long  = processor-in-the-loop
}
\DeclareAcronym{pp}{
    short = PP,
    long  = prioritized planning
}
\DeclareAcronym{qp}{
    short = QP,
    long  = quadratic programming,
}
\DeclareAcronym{rhgs}{
    short = RHGS,
    long  = receding horizon graph search,
    short-indefinite = an,
}
\DeclareAcronym{rhc}{
    short = RHC,
    long  = receding horizon control,
    short-indefinite = an,
}
\DeclareAcronym{rrt}{
    short = RRT,
    long  = rapidly-exploring random tree,
    short-indefinite = an,
}
\DeclareAcronym{rss}{
    short = RSS,
    long  = responsibility-sensitive safety,
    short-indefinite = an,
}
\DeclareAcronym{rti}{
    short = RTI,
    long  = real-time iteration,
    short-indefinite = an,
}
\DeclareAcronym{scdmpc}{
    short = SC-DMPC,
    long = Synchronization-Based Cooperative Distributed Model Predictive Control,
    short-indefinite = an
}
\DeclareAcronym{scp}{
    short = SCP,
    long  = sequential convex programming,
    short-indefinite = an,
}
\DeclareAcronym{scr}{
    short = SCR,
    long  = sequential convex restriction,
    short-indefinite = an,
}
\DeclareAcronym{sdo}{
    short = SDO,
    long  = saturation degree ordering,
    short-indefinite = an,
}
\DeclareAcronym{sgs}{
    short = SGS,
    long  = state-of-the-art graph search,
    short-indefinite = an,
}
\DeclareAcronym{sil}{
    short = SiL,
    long  = software-in-the-loop,
}
\DeclareAcronym{sl}{
    short = SL,
    long  = sequential linearization,
    short-indefinite = an,
}
\DeclareAcronym{sqp}{
    short = SQP,
    long  = sequential quadratic programming,
    short-indefinite = an,
}
\DeclareAcronym{tsp}{
    short = TSP,
    long  = traveling salesman problem,
}
\DeclareAcronym{uav}{
    short = UAV,
    long  = unmanned aerial vehicle,
    long-indefinite = an,
}
\DeclareAcronym{udlab}{
    short = IDS3C,
    long  = Information and Decision Science Scaled Smart City,
}
\DeclareAcronym{xil}{
    short = XiL,
    long  = X-in-the-loop,
    long-indefinite = an,
}

\newglossaryentry{def:agent}{
	name=agent,
	description={An agent is a system which is composed of at least one of the three elements: sensors, actuators, and a dynamic behavior.%
    },
}

\newglossaryentry{def:agentActive}{
	name=active agent,
	description={Active \glspl{def:agent} are \glspl{def:agent} which are connected using a communication
    network over which they can exchange data. The exchanged data is
    used by the \glspl{def:agent}' controllers to find appropriate inputs to achieve their
    goals while interacting with other \glspl{def:agent}.
    Additionally, active \glspl{def:agent} consider \glspl{def:agentPassive}},
    parent=def:agent,
}

\newglossaryentry{def:agentPassive}{
	name=passive agent,
	description={Passive \glspl{def:agent} are \glspl{def:agent} without networked control. However, they can communicate their data like current and future states to \glspl{def:agentActive}, or they can be detected by \glspl{def:agentActive}' sensors.%
    },
    parent=def:agent,
}

\newglossaryentry{def:distrutedSolutionQuality}{
	name=distributed solution quality,
	description={%
        The quality $q\in [0,1]$ of the solution in \ac{dmpc} is the networked objective function value ${\fcnObjective}_{c}$ for the solution of the corresponding \ac{cmpc} formulation divided by the objective function value ${\fcnObjective_d}$ for the solution of the \ac{dmpc} formulation
        \begin{equation}
            q = \frac{\fcnObjective_c}{\fcnObjective_d}.
        \end{equation}
    },
}

\newglossaryentry{def:mas}{
	name=multi-agent system,
	description={A system consisting of multiple \glspl{def:agent}.%
    },
}

\newglossaryentry{def:ncs}{
	name=networked control system,
	description={A system consisting of multiple \glspl{def:agentActive}.%
    },
}

\newglossaryentry{def:prediction}{
	name=prediction,
	description={
        A prediction $\agentPrediction\ofAgent{\anAgent}$ of \gls{def:agent} $\anAgent$ is its predicted state trajectory as obtained from the solution of its \ac{ocp} at time $\timestep$.
        A prediction $\agentPrediction\ofAgentInAgent{\anAgent}{\anotherAgent}$ of \gls{def:agent} $\anAgent$ for \gls{def:agent} $\anotherAgent$ is agent $\anotherAgent$'s state trajectory as viewed from agent $\anAgent$ at time $\timestep$. It is obtained by communication or by predicting \gls{def:agent} $\anotherAgent$'s state trajectory with its model using the solution to its \ac{ocp}.%
    },
}

\newglossaryentry{def:consistency}{
	name=prediction consistency,
	description={%
        \Iac{ncs} is prediction consistent at time step $\timestep$ if the \gls{def:prediction} $\agentPrediction\ofAgentInAgent{\anotherAgent}{\anAgent}$
            of all neighbors $\anotherAgent \in \setNeighbors{\anAgent}$
            in every agent $\anAgent\in\setAgents$
            equals their own \gls{def:prediction} $\agentPrediction\ofAgent{\anotherAgent}$, i.e.,
        \begin{equation}
            \agentPrediction\ofAgentInAgent{\anotherAgent}{\anAgent}=\agentPrediction\ofAgent{\anotherAgent}, \quad \forall \anAgent \in \setAgents, \forall \anotherAgent \in \setNeighbors{\anAgent}.
        \end{equation}%
    }
}

\newglossaryentry{def:ncsFeasible}{
	name=NCS-feasible,
	description={%
        The solutions $\sysControlInputs_{\cdot \vert \timestep}\ofAgent{\anAgent}$ of all agents $i\in\setAgents$ are \acs*{ncs}-feasible if the stacked solution vector $\bm{U}_{\cdot \vert \timestep} = \left( \sysControlInputs_{\cdot \vert \timestep}\ofAgent{1}, \ldots, \sysControlInputs_{\cdot \vert \timestep}\ofAgent{\numAgents} \right)\transposed$ satisfies all constraints of the corresponding central \acf*{ocp} considering all agents.%
    },
}

\newglossaryentry{def:feasibleAgent}{
	name=agent-feasible,
	description={%
        A solution is agent-feasible if it satisfies the constraints of to the corresponding agent's \ac{ocp}.%
    },
}

\newglossaryentry{def:networkedObjectiveFunction}{
	name=networked objective function,
	description={%
        The objective function value ${\fcnObjective}$ in \iac{ncs} formulation is the sum of all agent objective functions \fcnObjective[i]
        \begin{equation}
            \fcnObjective = \sum_{i}^{i\in\setAgents} \fcnObjective[i].
        \end{equation}
    },
}

\newglossaryentry{def:optimalPrioritization}{
	name=optimal prioritization,
	description={%
        The optimal prioritization results in the solution for every agent with the lowest networked objective function value obtainable by prioritization.%
    },
}

\newglossaryentry{def:graph_undirected}{
	name=undirected graph,
	description={%
        An undirected graph $\graphUndirected = \left(\setVertices,\setEdges\right)$ is a pair of two sets,
        the set of vertices $\setVertices=\set{1,\dots,\numAgents}$
        and the set of undirected edges $\setEdges \subseteq \setVertices \times \setVertices$.
        The edge between $i$ and $j$ is denoted by $\edgeUndirected{i}{j}$.
    },
}

\newglossaryentry{def:graph_directed}{
	name=directed graph,
	description={%
        A directed graph $\graphDirected = \left(\setVertices,\setEdgesDirected\right)$ is a pair of two sets,
        the set of vertices $\setVertices=\set{1,\dots,\numAgents}$
        and the set of directed edges $\setEdgesDirected \subseteq \setVertices \times \setVertices$.
        The edge from $i$ to $j$ is denoted by $\edgeDirected{i}{j}$.
        An oriented graph is a directed graph obtained from an undirected graph by replacing each edge $\edgeUndirected{i}{j}$ with either $\edgeDirected{i}{j}$ or $\edgeDirected{j}{i}$.
    },
}

\newglossaryentry{def:path}{
	name=path,
	description={%
        A path in a graph $\graphUndirected$ is a sequence of edges connecting distinct vertices
        \begin{equation}
            (e_z)_{z=1}^{n}
            \bigl(
                \edgeDirected{\anAgent_1}{\anAgent_2},
                \edgeDirected{\anAgent_2}{\anAgent_3},
                \ldots,
                \edgeDirected{\anAgent_{n}}{\anAgent_{n+1}}
            \bigr)
            ,
        \end{equation}
        with the domain $\setNaturalNumbers$ and the codomain $\setEdges$.
        The length of the path is defined as the number of edges $n$.
    },
}

\newglossaryentry{def:graph:adjacency}{
	name=adjacency,
	description={%
    A vertex $j$ is a predecessor of vertex $i$ iff $\edgeDirected{j}{i}\in\setEdges$.
    The set of predecessors of vertex $i$ is denoted by
    \begin{equation}
        \setPredecessors{i}=\set{j \mid \edgeDirected{j}{i}\in\setEdges}.
    \end{equation}
    A vertex $j$ is a successor of vertex $i$ iff $\edgeDirected{i}{j}\in\setEdges$.
    The set of successors of vertex $i$ is denoted by
    \begin{equation}
        \setSuccessors{i}=\set{j \mid \edgeDirected{i}{j}\in\setEdges}.
    \end{equation}
    A vertex $j$ is a neighbor to or adjacent to vertex $i$ if it is either a predecessor or a successor.
    The set of neighbors of vertex $i$ is denoted by
    \begin{equation}
        \setNeighbors{i}= \setSuccessors{i} \cup \setPredecessors{i}.
    \end{equation}
    },
}

\newglossaryentry{def:graph:degree}{
	name=degree,
	description={%
        The degree
        \begin{equation}
            \vertexDegree{i} = \lvert \setNeighbors{i} \rvert
        \end{equation}
        of a vertex $i$ denotes the number of its adjacent vertices.
        The number of incoming edges called in-degree is denoted by
        \begin{equation}
            \vertexInDegree{i} = \lvert \setPredecessors{i} \rvert.
        \end{equation}
        The number of outgoing edges called out-degree is denoted by
        \begin{equation}
            \vertexOutDegree{i} = \lvert \setSuccessors{i} \rvert.
        \end{equation}
    },
}

\newglossaryentry{def:graph:diameter}{
	name=diameter,
	description={%
        The diameter of a graph is the greatest length of any shortest path between each pair of vertices $\setVertices \times \setVertices$.%
    },
}

\newglossaryentry{def:coupling_graph_undirected}{
	name=undirected coupling graph,
	description={%
        An undirected coupling graph
        $\graphUndirected(\timestep)=\bigl(\setVertices,\setEdges(\timestep)\bigr)$
        is a graph that represents the interaction between agents at time step $\timestep$.
        Vertices $\anAgent\in\setVertices=\set{1,\ldots,\numAgents}$ represent agents,
        and edges $\edgeUndirected{\anAgent}{\anotherAgent}\in\setEdges$ represent coupling objectives and constraints between agents.
    },
}

\newglossaryentry{def:coupling_graph_directed}{
	name=directed coupling graph,
	description={%
        A directed coupling graph
        $\graphDirected(\timestep)=\bigl(\setVertices,\setEdgesDirected(\timestep)\bigr)$
        is an oriented graph obtained from orienting the edges of an undirected coupling graph at time step $\timestep$.
        If an edge $\edgeDirected{\anAgent}{\anotherAgent}$ is directed from agent $\anAgent$ to agent $\anotherAgent$, then agent $\anotherAgent$ is responsible for considering the respective coupling objective and constraint in its planning problem.
    },
}

\newglossaryentry{def:matrix:Adjacency}{
	name=adjacency matrix,
	description={An adjacency matrix represents a graph with $\numAgents$ vertices in a matrix $\matAdjacency \in \set{0,1}^{\numAgents\times\numAgents}$ with entries
    \begin{equation}
        \matAdjacencyElement{ij} =
            \begin{cases}
                1 & \text{ if } \edgeDirected{i}{j} \in \setEdges \\
                0 & \text{ otherwise.}
            \end{cases}
    \end{equation}
    },
}

\newglossaryentry{def:tCompNcs}{
	name=computation time of \iac{ncs},
	description={%
        The computation time $\tCompNcs$ of \iac{ncs} is the time required for the \ac{ncs} to measure the states, formulate and solve the \ac{ocp}, apply the inputs to all agents, and communicate the required data. 
    },
}

\newglossaryentry{def:setReachable}{
	name=reachable set,
	description={%
        The reachable set of states $\setReachable$ of an agent from an initial time $t_{\text{init.}}$ to an end time $t_{\text{end}}$ is
            \begin{equation}\label{eq:setReachable}
                \setReachable_{[t_{\text{init.}},t_{\text{end}}] \mid t_{\text{init.}}} = \biggl\{ \int_{t_{\text{init.}}}^{t_{\text{end}}} \sysModelContinuous(\sysState,\sysControlInputs)dt
                \biggm| \sysState(t_{\text{init.}}) \in \setFeasibleStates(t_{\text{init.}}), \forall t: \sysControlInputs \in \setFeasibleInputs \biggr\},
            \end{equation}
        with the possible system initial states $\sysState(t_{\text{init.}})$ being bounded by its initially admissible set $\setFeasibleStates(t_{\text{init.}}) \subseteq \setRealNumbers^{\numStates}$, and the possible system control inputs $\sysControlInputs$ being bounded by its admissible set $\setFeasibleInputs \subseteq \setRealNumbers^{\numInputs}$.
    },
}

\newglossaryentry{def:conflictualDecisions}{
	name=conflictual decisions in \iac{ncs},
	description={%
        Consider two decisions made by two agents of \iac{ncs} at time step $\timestep$ with a duration $N_k$.
        They are deemed conflictual if the predicted outcome of the decisions violates the \ac{ncs}-feasibility at some point in time.%
    },
}

\newglossaryentry{def:conflictualSpace}{
	name=conflictual space of \iac{ncs},
	description={%
        In dynamic systems, the state space represents the set of all possible states the systems can occupy. 
        The conflictual space refers to a subset, or potentially the entirety, of this state space where whether decisions are conflictual is determined.
    },
}

\newglossaryentry{def:anytimePlanner}{
	name=Anytime trajectory planner,
	description={%
        An anytime trajectory planner is a trajectory planner that quickly identifies a feasible trajectory and incrementally improves it over time.
    },
}

\definecolor{color1}{RGB}{0, 84, 159}      %
\definecolor{color2}{RGB}{246, 168, 0}     %
\definecolor{color3}{RGB}{0, 152, 161}     %
\definecolor{color4}{RGB}{122, 111, 172}   %
\definecolor{color5}{RGB}{204, 7, 30}      %
\definecolor{color6}{RGB}{0, 97, 101}      %
\definecolor{color7}{RGB}{189, 205, 0}     %
\definecolor{color8}{RGB}{161, 16, 53}     %
\definecolor{color9}{RGB}{87, 171, 39}     %
\definecolor{color10}{RGB}{97, 33, 88}     %
\definecolor{color11}{RGB}{255, 237, 0}    %

\newcommand{\footremember}[2]{%
    \footnote{#2}
    \newcounter{#1}
    \setcounter{#1}{\value{footnote}}%
}
\newcommand{\footrecall}[1]{%
    \footnotemark[\value{#1}]%
}

\begin{document}

    \title{Graph Coloring to Reduce Computation Time in Prioritized Planning}

    \author{%
        Patrick Scheffe\,\orcidlink{0000-0002-2707-198X} \footremember{rwth}{Chair of Embedded Software, RWTH Aachen University,
        Im Süsterfeld 9, 52072 Aachen, Germany}%
        \and Julius Kahle\,\orcidlink{0000-0003-3986-1986} \footrecall{rwth}%
        \and Bassam Alrifaee\,\orcidlink{0000-0002-5982-021X} \footremember{unibw}{Department of Aerospace Engineering, University of the Bundeswehr Munich,
        Werner-Heisenberg-Weg 39, 85579 Neubiberg, Germany}%
    }
    \date{}

\maketitle

\begin{abstract}
    Distributing computations among agents in large networks reduces computational effort in \ac{mapf}.
    One distribution strategy is \ac{pp}.
    In \ac{pp}, we couple and prioritize interacting agents to achieve a desired behavior across all agents in the network.
    We characterize the interaction with \iac{dag}.
    The computation time for solving \ac{mapf} problem using \ac{pp} is mainly determined through the longest path in this \ac{dag}.
    The longest path depends on the fixed undirected coupling graph and the variable prioritization.
    The approaches from literature to prioritize agents are numerous and pursue various goals.
    This article presents an approach for prioritization in \ac{pp} to reduce the longest path length in the coupling \ac{dag} and thus the computation time for \ac{mapf} using \ac{pp}.
    We prove that this problem can be mapped to a graph-coloring problem, in which the number of colors required corresponds to the longest path length in the coupling \ac{dag}.
    We propose a decentralized graph-coloring algorithm to determine priorities for the agents.
    We evaluate the approach by applying it to \ac{mamp} for \acp{cav} on roads using, a variant of \ac{mapf}.
\end{abstract}

\acresetall

\medskip
\noindent
\begin{tabular}{@{} l @{ } l @{}}
    \textbf{Code}  & \href{https://github.com/embedded-software-laboratory/p-dmpc}{\small github.com/embedded-software-laboratory/p-dmpc}
\end{tabular}

\newcommand{\graphMap}      {\ensuremath{\graphUndirected_M}}
\newcommand{\setVerticesMap}{\ensuremath{\setVertices_M}}
\newcommand{\setEdgesMap}   {\ensuremath{\setEdges_M}}
\newcommand{\aVertex}       {\ensuremath{v}}
\newcommand{\anotherVertex} {\ensuremath{u}}

\section{Introduction}
\label{sec:introduction}
\subsection{Motivation}
In multi-agent planning problems, several agents may share a common objective or must respect common constraints.
Traditional \ac{mapf} is a problem where multiple agents navigate through an environment modeled as a graph $\graphMap=(\setVerticesMap,\setEdgesMap)$.
This graph consists of a set of vertices $\setVerticesMap$, representing locations, and a set of edges $\setEdgesMap$, representing the paths connecting these locations.
Each agent $\anAgent$ aims to move from its starting vertex $s_{\anAgent} \in \setVerticesMap$ to its target vertex $t_{\anAgent} \in \setVerticesMap$.
The primary constraint for each agent is to avoid collisions with other agents.
A collision occurs if two agents, $\anAgent$ and $\anotherAgent$, occupy the same vertex $\aVertex$ at the same time step $\timestep$, which can be represented by the tuple $\tuple{\anAgent, \anotherAgent, \aVertex, \timestep}$.
Additionally, a collision can also happen when both agents traverse the same edge $\edgeUndirected{\aVertex}{\anotherVertex}\in\setEdgesMap$, denoted by the tuple $\tuple{\anAgent, \anotherAgent, \aVertex, \anotherVertex, \timestep}$.

Solving \iac{mapf} problem optimally is NP-hard \citep{yu2013structure,yu2016intractability}.
Centralized approaches exploit the problem structure or lazily explore the solution space in order to increase their computational efficiency \citep[e.g.,][]{sharon2015conflictbased,yu2020average,pan2024rhecbs,li2021eecbs,barer2021suboptimal,guo2024expected}. %
\Ac{pp}, introduced by \cite{erdmann1987multiple}, is a computationally more efficient method for \ac{mapf}.
Instead of solving \iac{mapf} problem optimally, \ac{pp} divides the problem into multiple subproblems, and each agent solves only its subproblem.
Since the subproblems are smaller in size compared to the centralized problem, the computational complexity is reduced.
Additionally, agents can compute the solution in a distributed fashion.

Since only a subset of agents is considered in each subproblem, a method is necessary to find consensus among agents.
Consensus, in general, can be established, e.g., by iteration \citep{trodden2013cooperative,kloock2020prediction,blasi2018Distributed,kloock2023coordinated,cap2015prioritized}.
In \ac{pp}, this is solved by each agent bearing a priority.
Agents with lower priority are obligated to avoid collision with agents bearing higher priority.
This requires each agent to consider the higher priority agents as dynamic obstacles while planning.
After solving the subproblem, the resulting plans are exchanged among agents and considered in the planning problems of lower priority agents.
This requires sequential planning in order of decreasing priority \citep{vandenberg2005prioritized}.
Thus, the maximum computation time of \ac{pp} for \ac{mapf} scales approximately linearly with the number of agents.

The feasibility, optimality, and computation time of a solution in \ac{pp} depends on the prioritization, among other factors.
This poses a challenge in large-scale \ac{mapf} problems when computation time is limited.

\subsection{Related Work}
In \ac{pp}, an agent $i$ solves its \ac{mapf} subproblem by optimizing its own solution considering the communicated predictions of higher-priority agents as dynamic obstacles.
\begin{definition}[Prediction]\label{def:prediction}
    A prediction $\agentPrediction\ofAgent{\anotherAgent}$ of agent $\anotherAgent$
        is \gls{def:agent} $\anotherAgent$'s optimized solution at time $\timestep$.
    A prediction $\agentPrediction\ofAgentInAgent{\anotherAgent}{\anAgent}$ of agent $\anotherAgent$ in agent $\anAgent$
        is \gls{def:agent} $\anotherAgent$'s prediction from agent $\anAgent$'s perspective at time $\timestep$.
\end{definition}

One of the difficulties in distributed optimization approaches such as \ac{pp} is guaranteeing prediction consistency among the agents in the network \citep{alrifaee2016coordinated, kloock2020prediction,trodden2013cooperative}, or compensating for the lack thereof.
Let the set of agent $i$'s neighbors be denoted by $\setNeighbors{i}$ and equal the set of agents that agent $i$ considers in its subproblem (see \cref{sec:introduction:notation} for a precise definition).
\begin{definition}[Prediction Consistency]\label{def:consistency}
    A network is prediction consistent at time step $\timestep$ if the \gls{def:prediction} $\agentPrediction\ofAgentInAgent{\anotherAgent}{\anAgent}$
            of all neighbors $\anotherAgent \in \setNeighbors{\anAgent}$
            in every agent $\anAgent\in\setAgents$
            equals their own \gls{def:prediction} $\agentPrediction\ofAgent{\anotherAgent}$, i.e.,
        \begin{equation}
            \agentPrediction\ofAgentInAgent{\anotherAgent}{\anAgent}=\agentPrediction\ofAgent{\anotherAgent}, \quad \forall \anAgent \in \setAgents, \forall \anotherAgent \in \setNeighbors{\anAgent}.
        \end{equation}%
\end{definition}

We differentiate between parallel and sequential \ac{pp}.
Parallel \ac{pp} has nearly constant computation time, but can lead to a loss of prediction consistency.
In contrast, sequential \ac{pp} guarantees prediction consistency, but the computation time of the \ac{mapf} instance increases linearly with the number of agents.

\subsubsection{Parallel PP}
In parallel \ac{pp}, agents compute solutions parallelly and exchange predictions afterwards.
Assuming no delay introduced by the communication, the communicated predictions can be used no earlier than in the following time step.
Each agent consequently shifts the received predictions by one time step such that the prediction time matches the time in its planning problem.
Assuming continuous computation, the entries at the end of the prediction are missing after such a shift and must be estimated \citep{alrifaee2017networked}.
When an agent deviates from the prediction it communicated in the previous time step, the predictions become inconsistent.
Since the subproblems are then solved using different data, the \ac{mapf} solution might inadvertently violate constraints or become unstable.
Consequently, parallelizing the computation requires dealing with the problem of prediction inconsistency, as presented in the following literature review.

A solution to the problem of prediction inconsistency is to constrain an agent's change from a previous prediction.
\cite{dunbar2012distributed} and \cite{zheng2017distributed} show stability of \ac{pp} with parallel computation applied to vehicle platoons.
Nearly consistent predictions are achieved with a penalty on changes of previously predicted plans and a terminal constraint.
\cite{dunbar2006distributed} stabilize a distributed formation controller by limiting the control action to restrain deviations from previously predicted plans.
Another approach to enable parallel computation in linear time-invariant systems is tube-based \ac{dmpc}, in which each controller rejects bounded disturbances \citep{richards2007robust}.
While there are improvements to tighten the tubes \citep{farina2012distributed, riverso2012tubebased}, a downside to tube-based \ac{mpc} remains the conservativeness of its solutions.
In our previous work \citep{scheffe2024limiting}, the predictions of neighbors, and thus the constraints, are overapproximated by their reachable set.
Consequently, agents can compute in parallel while guaranteeing satisfaction of the original constraint.
However, the overapproximation tightens the constraints compared to sequential \ac{pp} and therefore leads to more conservative solutions.

Although successful in their respective applications, the above approaches have higher input constraints and therefore less flexibility when solving the planning problem.
This might lead to more conservative solutions or even constraint violations.

\subsubsection{Sequential PP}
In sequential \ac{pp}, agents compute their solutions sequentially.
An agent waits for the prediction of higher prioritized agents before it solves its own planning problem.
In one time step, every agent in the network agrees on the predictions of others.
Thus, sequential \ac{pp} guarantees prediction consistency among the agents in the network \citep{alrifaee2016coordinated}, but suffers from increasing computation time.

The following literature review considers approaches in which agents compute sequentially.
In sequential computation of \ac{pp}, the prioritization determines the computation order of agents besides the aforementioned responsibility to consider higher prioritized agents in their subproblem.
There are $\numAgents!$ prioritizations, with $\numAgents \in \setNaturalNumbers$ being the number of agents in the network.
In large-scale networks, the number of prioritizations is too large to evaluate online.
To reduce the computational effort required to find the best prioritization, several heuristics have been proposed.

The following heuristics pursue the goal of achieving feasible solutions for every agent in the network.
\cite{wu2020multirobot} prioritizes agents that have fewer options to achieve their goal.
In our previous work \citep{kloock2019distributed}, multiple car-like robots simultaneously move from an arbitrary start pose to a defined target pose in a confined space.
We prioritize robots to increase the feasibility of this go-to-formation problem: The more obstructed the target pose in the formation is to reach for a robot, the higher its priority.
The highest priority is given to robots that need to stop in front of a space boundary, the second highest to robots that need to stop in front of another robot.
The lowest priority is given to robots with a target pose in free space, next to another robot or next to a space boundary\footnote{\url{https://www.youtube.com/watch?v=XGql8FrjW6I}}.
\cite{luo2016distributed} and \cite{scheffe2022increasing} let agents plan iteratively and communicate an optimal solution without considering other agents.
The agents receive a priority based on the number of predicted collisions with the plans of other agents.
This also resembles the idea of prioritizing agents that are more restricted.

The following are objective-based heuristics and prioritize agents to improve the quality of the \ac{mapf} solution.
In our previous work \citep{kloock2019distributeda}, we prioritize agents with the goal of increasing the traffic flow rate at a road intersection.
The objective function penalizes a deviation from a reference speed and a change of acceleration.
In order to reduce changes in the acceleration and therefore increase the networked solution quality, we prioritize agents if they are faster and closer to the intersection.
The less time a vehicle has left before it needs to start decelerating to come to a full stop in front of the intersection, the higher its priority.
In the work of \cite{vandenberg2005prioritized}, robots with longer estimated travel distances receive higher priority to more evenly distribute the travel time among robots.
The algorithm presented by \cite{chalaki2022priorityaware} prioritizes vehicles at intersections using job-shop scheduling.
\cite{bennewitz2002finding} prioritizes agents are randomly.
Priorities are swapped iteratively to increase the quality of the \ac{mapf} solution.
The planning problem has to be solved in each iteration, which increases computation effort.
\cite{zhang2022learning} present a prioritization algorithm based on machine learning that performs competitively compared to heuristic algorithms.

Many of these approaches affect the feasibility or the optimality of the agents' solutions to the \ac{mapf} problem.
However, they do not consider the influence of the prioritization on the computation time to solve \ac{mapf} using \ac{pp}.

\subsubsection{Parallelization in Sequential PP}
When agents need to consider common objectives or constraints, we term these agents coupled.
For agents that are not coupled, prioritizations exist such that their computations can be parallelized without risking an inadvertent violation of constraints \citep{alrifaee2016coordinated,alrifaee2017networked}.
The goal of minimizing the number of sequential computations corresponds to a timetabling problem.
In the work of \cite{welsh1967upper}, an incompatibility matrix represents the coupling of jobs.
The problem is to find the minimum time needed to perform the jobs, which is solved by graph coloring.
Jobs with the same color can be scheduled in the same time slot.

This concept is transferred to \ac{pp} for distributed trajectory planning of vehicles by \cite{kuwata2007distributed}.
A coupling graph connects vertices with an edge if the corresponding vehicles are within a certain distance.
Their work proves that vehicles without edges can plan their trajectories simultaneously.
To maximize the number of simultaneous computations, a central entity determines the planning order through a heuristic coloring algorithm.
All vehicles associated with vertices of the same color can compute simultaneously.
\citep{kuwata2007distributed} do not prove that prioritization through graph coloring is equivalent to maximizing the number of simultaneous computations in \ac{pp}.
Additionally, the centralized entity which performs graph coloring introduces high effort of communication and represents a single point of failure for the system.

\subsection{Contribution of this Article}
In \ac{pp}, the number of agents whose computations can be parallelized depends on the coupling and the prioritization.
The coupling is often determined by the application.
Therefore, the computation time to solve \ac{mapf} using \ac{pp} can only be influenced through the prioritization.
We present a prioritization algorithm to reduce computation time by maximizing the number of parallel agent computations.
We formalize the problem of maximizing the number of parallel agent computations and prove the problem's equivalence to a graph coloring problem.
We formulate our algorithm for graph coloring such that it can be solved in a decentralized fashion by each agent.

\subsection{Notation}\label{sec:introduction:notation}
In this paper, we speak of agents whenever concepts are generally applicable to \ac{pp}.
A variable $x$ is marked with a superscript $x^{(\anAgent)}$ if it belongs to agent $\anAgent$.
The actual value of a variable $x$ at time $\timestep$ is written as $x_{\timestep}$, while values predicted for time $\timestep+\timestepIterator$ at time $\timestep$ are written as $x_{\timestep+\timestepIterator \vert \timestep}$.
A trajectory is denoted by replacing the time argument with $(\cdot)$ as in $x_{\cdot \vert \timestep}$.
For any set $\mathcal{S}$, the cardinality of the set is denoted by $|\mathcal{S}|$. 

We use graphs as a modeling tool of networks.
Every agent is associated with a vertex, so the terms are used synonymously.
\begin{definition}[\Gls{def:graph_directed}]\label{def:graph}
    \glsdesc*{def:graph_directed}
\end{definition}
We characterize the relation between agents by their adjacency.
\begin{definition}[\Gls{def:graph:adjacency}]\label{def:graph:adjacency}
    \glsdesc*{def:graph:adjacency}
\end{definition}
We assume that states of agents are observable, and therefore we refer to states instead of outputs.
The definitions and methods can be transferred to outputs as well, which we omit for brevity.

\subsection{Structure of this Article}
The rest of this article is organized as follows.
First, our \ac{pp} framework for \ac{mamp} is introduced in \cref{sec:pp}.
\Ac{mamp} is a variant of \ac{mapf}, in which the kinodynamic constraints of agents are taken into account during planning.
The problem that the present paper addresses is formally defined in \cref{sec:problem_statement}, before our solution to it using graph coloring is presented in \cref{sec:problem_solution}.
We show numerical results of our approach in \cref{sec:evaluation} by applying it to \ac{mamp} for \acp{cav} at intersections.

\section{Prioritized Planning for Multi-Agent Motion Planning}
\label{sec:pp}
This section provides a background to \ac{pp} applied to \ac{mamp} and prioritization.
\Ac{mamp} is a variant of \ac{mapf} which can be applied to \acp{cav}.
While \ac{mapf} abstracts the environment and system dynamics in a graph representation, \ac{mamp} explicitly considers the system dynamics as \acp{ode} and models the environment by static obstacles represented as polygons.
Further, \ac{mapf} aims at planning a complete path from a start to a goal vertex.
Contrary, the objective in our \ac{mamp} application is to plan for a fixed time horizon and to shift the horizon every time step, commonly known as receding horizon planning or online replanning \citep{cashmore2019replanning,silver2005cooperative,li2021lifelong,scheffe2023receding,shahar2021safe}.
The planning problem in \ac{mamp} is modeled as \iac{ocp}.
Certainly, there exists work that extend beyond traditional categorization of \ac{mapf} by, e.g., incorporating system dynamics \citep{alonso-mora2018cooperative,luo2016distributed,le2018cooperative}.

\begin{figure}
    \centering
    \includegraphics{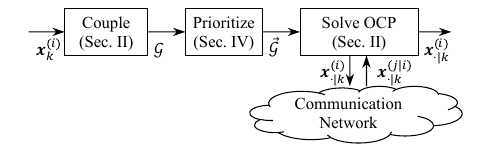}
    \caption{
        Our framework of \ac{pp} which runs every time step.
        $\sysStateAgentTime{i}{k}$ is the state,
        $\graphUndirected$ is the undirected coupling graph,
        $\graphDirected$ is the directed coupling graph,
        $\agentPrediction\ofAgent{i}$ is the prediction of agent $i$,
        $\agentPrediction\ofAgentInAgent{j}{i}$ is the the prediction of agent $\anotherAgent$ in agent $\anAgent$.
        }
    \label{fig:pp_framework}
\end{figure}

Our framework \ac{pp} framework for \ac{mamp} is illustrated in \cref{fig:pp_framework}.
Agents solve their \ac{ocp} (\cref{sec:pp:solve-ocp}) when they have received all predictions from their predecessors, and send their predictions to their successors.
To determine the predecessors and successors of agents, we couple (\cref{sec:pp:coupling}) and prioritize (\cref{sec:problem_solution}) them.
This paper focuses on the prioritization step, which we evaluate in the application of \ac{mamp} for \acp{cav}.
We sketch coupling and solving the \ac{ocp} for trajectory planning in the following for context.
Whenever we refer to a \ac{mamp} problem incorporating the entire network of agents, we term this as an \ac{mamp} instance.

\subsection{Couple}
\label{sec:pp:coupling}
If agents interact via their objective or constraint functions, we speak of coupled agents. A coupling graph represents the interaction between agents.
\begin{definition}[\Gls{def:coupling_graph_directed}]\label{def:coupling-graph}
    \glsdesc*{def:coupling_graph_directed}
\end{definition}
The application in which \ac{mamp} is used determines which coupling objectives and constraints must be considered in the \acp{ocp}.
For example, collision freeness between robots can be achieved via coupling constraints.

The undirected coupling graph can often be determined in networked applications before solving the \ac{ocp}.
In robot applications, if robots move on predetermined paths, we can couple robots if these paths intersect.
Paths can be predetermined for road vehicles following lanes, or for warehouse robots \citep{ma2017lifelong}.
In \ac{mamp}, agents move freely.
We analyze their reachable set for the prediction horizon, and couple them if their reachable sets intersect, similar to \cite{scheffe2024limiting}.

\subsection{Solve OCP}
\label{sec:pp:solve-ocp}
The objective of an agent $\anAgent$ is to follow a reference trajectory:
\begin{equation}\label{eq:objective}
    \fcnObjective[\anAgent](k) =
    \sum_{\timestepIterator=1}^{\horizonPrediction} \fcnObjectiveState[\anAgent] \left( \sysState_{\timestep+\timestepIterator | \timestep}^{(\anAgent)},\bm{r}_{\timestep+\timestepIterator | \timestep}^{(\anAgent)} \right).
\end{equation}
$\horizonPrediction\in\setNaturalNumbers$ is the prediction horizon, $\numStates\in\setNaturalNumbers$ is the number of states, and $\numInputs\in\setNaturalNumbers$ is the number of inputs.
The function $ \fcnObjectiveState[\anAgent] \colon \setRealNumbers^{\numStates}\times\setRealNumbers^{\numStates}\to\setRealNumbers$ penalizes a deviation to the reference trajectory $\bm{r}_{\cdot | k}^{(i)}$ of agent $i$ and constitutes the objective function \cref{eq:pdmpc:ocp_obj} in the following \ac{ocp}.
The \ac{ocp} is solved inside the \ac{mamp} of each agent $\anAgent\in\setVertices$ at each time step $k$:
\begin{mini!}
    {
        \sysControlInputs\ofAgent{\anAgent}\forTimeAtTime{\cdot}{\timestep}
    }
    {
        \fcnObjective\ofAgent{\anAgent}(k)
        \label{eq:pdmpc:ocp_obj}
    }
    {
        \label{eq:4-ocp}
    }
    {
    }
    \addConstraint{
        \sysState\ofAgent{\anAgent}\forTimeAtTime{\timestep+\timestepIterator+1}{\timestep}
    }
    {
        = \sysModelDiscrete \big(
            \sysState\ofAgent{\anAgent}\forTimeAtTime{\timestep+\timestepIterator}{\timestep},
            \sysControlInputs\ofAgent{\anAgent}\forTimeAtTime{\timestep+\timestepIterator}{\timestep}
        \big)
        ,\quad
        \protect\label{eq:pdmpc:ocp_c_system}
    }
    {
        \timestepIterator=0,\ldots,\horizonPrediction-1
    }
    \addConstraint{
        \sysState\ofAgent{\anAgent}\forTimeAtTime{\timestep+\timestepIterator}{\timestep}
    }
    {
        \in \setFeasibleStates
        ,\quad
        \protect\label{eq:pdmpc:ocp_c_state}
    }
    {
        \timestepIterator=1,\ldots,\horizonPrediction-1
    }
    \addConstraint{
        \sysState\ofAgent{\anAgent}\forTimeAtTime{\timestep+\horizonPrediction}{\timestep}
    }
    {
        \in \setFeasibleStates_f
        \protect\label{eq:pdmpc:ocp_c_final_state}
    }
    \addConstraint{
        \sysState\ofAgent{\anAgent}\forTimeAtTime{\timestep}{\timestep}
    }
    {
        = \sysState\ofAgent{\anAgent}(\timestep)
        \protect\label{eq:pdmpc:ocp_c_init_state}
    }
    \addConstraint{
        \sysControlInputs\ofAgent{\anAgent}\forTimeAtTime{\timestep+\timestepIterator}{\timestep}
    }
    {
        \in \setFeasibleInputs
        ,\quad
        \protect\label{eq:pdmpc:ocp_c_input}
    }
    {
        \timestepIterator=0,\ldots,\horizonPrediction-1
    }
    \addConstraint{
        \fcnConstraintCoupling\ofAgent{\anAgent \seqCoupling \anotherAgent} \left(
            \sysState\ofAgent{\anAgent}\forTimeAtTime{\timestep+\timestepIterator}{\timestep},
            \sysState\ofAgent{\anotherAgent}\forTimeAtTime{\timestep+\timestepIterator}{\timestep}
        \right)
    }
    {
        \leq 0
        ,\quad
        \protect\label{eq:pdmpc:ocp_c_coupling_sequential}
    }
    {
        \timestepIterator = 1,\ldots, \horizonPrediction, \quad
        \anotherAgent \in \setPredecessors{\anAgent}(\timestep)
        .
    }
\end{mini!}
The vector field $\sysModelDiscrete \colon \setRealNumbers^{\numStates}\times\setRealNumbers^{\numInputs}\to\setRealNumbers^{\numStates}$ in \cref{eq:pdmpc:ocp_c_system} resembles the discrete-time system model with $\numStates\in\setNaturalNumbers$ as the number of states and $\numInputs\in\setNaturalNumbers$ as the number of inputs.
The input to the model $\sysModelDiscrete$ consists of a vector $\sysControlInputs_{\cdot | \timestep} \in \setFeasibleInputs$ which is the control input and a vector $\sysState_{\cdot} \in\setFeasibleStates$ which is the agent's system state.
The set of feasible control inputs is denoted as $\setFeasibleInputs\subseteq\setRealNumbers^{\numInputs}$ and the set of feasible system states is denoted as $\setFeasibleStates\subseteq\setRealNumbers^{\numStates}$.
$\setFeasibleStates_f\subseteq\setFeasibleStates$ denotes the set of feasible terminal states.
This allows to further constrain this system state to achieve stable solutions.
The function $\fcnConstraintCoupling\ofAgent{\anAgent \seqCoupling \anotherAgent} \colon \setRealNumbers^{\numStates} \times \setRealNumbers^{\numStates} \to \setRealNumbers$ in \cref{eq:pdmpc:ocp_c_coupling_sequential} resembles the coupling constraint with predecessors.
The set of predecessors $\setPredecessors{\anAgent}(\timestep)$ contains the neighbors of $\anAgent$ with higher priority, see \cref{sec:problem_solution}.

\section{Problem Statement}
\label{sec:problem_statement}
The goal of our prioritization function is to reduce the computation time of the \ac{mamp} instance.
Agents can compute decentralized components of the \ac{pp} framework in parallel, i.e., coupling and prioritizing in \cref{fig:pp_framework}.
However, coupled agents must plan in sequence to guarantee prediction consistency.
When all agents plan sequentially, the computation time grows approximately linearly with the number of agents.
However, if some agents are uncoupled, they can potentially plan in parallel without affecting the prediction consistency.
An agent starts to plan as soon as it has received the predictions from each of its predecessors.
This offers a chance to reduce the computation time.
We term the highest number of sequentially planning agents the number of computation levels $\numLevels$ in the \ac{mamp} instance.
The number of computation levels corresponds to the longest path in the directed coupling graph $\graphDirected$.
With the following definitions, we can formalize the number of computation levels.
\begin{definition}[\Gls{def:graph:degree}]\label{def:graph:degree}
    \glsdesc*{def:graph:degree}
\end{definition}
Let the sources of $\graphDirected$ be the set of vertices without incoming edges
\begin{equation}
    \setVertices_s = \set{i\in \setVertices \mid \vertexInDegree{i} = 0}
\end{equation}
and the sinks be the set of vertices without outgoing edges
\begin{equation}
    \setVertices_t = \set{i\in \setVertices \mid \vertexOutDegree{i} = 0}.
\end{equation}
Note that since the coupling graph is a \ac{dag}, there is at least one source and one sink.
The number of computation levels $\numLevels$ is the length of the longest path between $\setVertices_s$ and $\setVertices_t$ in $\graphDirected$.
Note that the number of computation levels depend on the coupling and the prioritization of the agents.

All components of our \ac{pp} framework considered, the computation time of an \ac{mamp} instance consists of the coupling time, the prioritization time, the planning time, and the communication time.
The computation time which is influenced most by prioritization is the prioritization time and the planning time.
We introduce the following assumptions to state the addressed problem.
\begin{assumption}\label{ass:equal_comp_time}
    The planning times $\tComp^{(i)}$ of all agents $i$ are similar,
    \begin{equation}
        \tComp^{(i)} \approx \tComp \quad \forall i \in \setVertices.
    \end{equation}
\end{assumption}
\Cref{ass:equal_comp_time} is mild if we use an anytime algorithm for planning.
An anytime algorithm aims to find an initial solution quickly and then incrementally improve the solution as time allows \citep{likhachev2003ara}.
Limiting the planning time results in nearly constant planning times of all agents.
\begin{assumption}\label{ass:comp_time_highest}
    The planning time of an agent is far greater than the prioritization time,
    \begin{equation}
        \tComp\ofAgent{\anAgent} \gg T\ofAgent{\anAgent}_{\text{prio}}
    \end{equation}
\end{assumption}
\Cref{ass:comp_time_highest} is reasonable if an involving planning problem is solved in the \ac{mamp}.
This is the case if, e.g., the \ac{ocp} solved in \ac{mamp} is nonconvex, as in trajectory planning for \acp{cav}.

According to \cref{ass:comp_time_highest}, we can overapproximate the prioritization time as the maximum of all prioritization times $T\ofAgent{\anAgent}_{\text{prio}}$ for simplicity.
To obtain the computation time $\tCompNcs$ of the \ac{mamp} instance, we weigh the vertices $\setVertices$ with their respective planning time by a weighting function $f_w \colon \setVertices \to \mathbb{R}$
\begin{equation}\label{eq:weight_func}
    f_w (i) = \tComp^{(i)},
\end{equation}
with the planning time $\tComp^{(i)}$ of agent $i$ required to solve its subproblem.
Let $P_{\max}\subseteq\setVertices$ denote the longest path between $\setVertices_s$ and $\setVertices_t$ in $\graphDirected$ weighted with $f_w$.
The sum of the maximum prioritization time and the planning times along this path corresponds to the computation time of the \ac{mamp} instance
\begin{equation}\label{eq:computation_time}
    \tCompNcs = \max_{\anAgent \in \setAgents} \left( T\ofAgent{\anAgent}_{\text{prio}} \right) + \sum_{i\in P_{\max}} \tComp^{(i)}.
\end{equation}

Under \cref{ass:equal_comp_time,ass:comp_time_highest}, the computation time of the \ac{mamp} instance is mainly influenced by the number of levels $\numLevels$ of a coupling \ac{dag}.
This coupling \ac{dag} is constructed from the undirected coupling graph with the prioritization function $\fcnPrio$.
Since our goal is to decrease computation time, we formally state our problem as follows.
\begin{problem}\label{prob:min_level}
Given an undirected coupling graph $\graphUndirected$, determine a prioritization function $\fcnPrio^*$ that minimizes the number of levels $\numLevels$:
\begin{equation}
    \fcnPrio^* = \argmin_{\fcnPrio} \numLevels (\graphUndirected, \fcnPrio).
\end{equation}
\end{problem}

\section{Problem Solution}
\label{sec:problem_solution}
This section first introduces the concept of prioritization in \ac{pp} in \cref{sec:problem_solution:priority}.
We prove the equivalence of \cref{prob:min_level} to a vertex coloring problem in \cref{sec:prso:coloring}.
We solve \cref{prob:min_level} with the decentralized algorithm presented in \cref{sec:prso:algo}.

\subsection{Priority Idea}
\label{sec:problem_solution:priority}
In \ac{pp}, we prioritize the agents in the network \citep{kuwata2007distributed,alrifaee2016coordinated}.
Prioritizing agents results in clear responsibilities and determines the order of computations in sequential \ac{pp}.
A prioritization function $\fcnPrio \colon \setVertices\to\mathbb{N}$ prioritizes every agent's vertex in the network.
If $\fcnPrio(i)<\fcnPrio(j)$, then agent $i$ has a higher priority than agent $j$, or agent $i$ is prioritized over agent $j$.

By prioritizing, we can transform an undirected coupling graph into a directed coupling graph which is \iac{dag}.
An edge points towards the vertex with lower priority,
\begin{equation}\label{eq:construction-rule}
    \edgeDirected{i}{j} \in \setEdgesDirected \iff \fcnPrio (i) < \fcnPrio(j) \land \edgeDirected{i}{j}\in \setEdges.
\end{equation}
For the orientation of every edge to be well-defined, a valid prioritization function needs to assign pairwise different priorities to vertices that are connected by an edge, i.e.,
\begin{equation}
    \label{eq:valid_priority}
    \fcnPrio(i)\neq \fcnPrio(j), \quad \forall i, j \in \setVertices, i \neq j, \edgeDirected{i}{j}\in\setEdges.
\end{equation}
\begin{lemma}
    \label{lemma:priority_dag}
    Given the construction rule in \cref{eq:construction-rule}, the directed coupling graph $\graphDirected$ resulting from the undirected coupling graph $\graphUndirected$ and a valid prioritization function $\fcnPrio$ regarding \cref{eq:valid_priority} is a \ac{dag}.
\end{lemma}
\begin{proof}
    We prove this by contradiction analogous to \cite{yuan2017effective}.
    Assume the directed graph $\graphDirected$ to contain a cycle consisting of the edges $\edgeDirected{i_1}{i_2},\dots,\edgeDirected{i_{n-1}}{{i_n}},\edgeDirected{i_n}{i_1}$.
    From the construction rule \cref{eq:construction-rule} of $\graphDirected$ and from the transitive property of the ``$<$'' relation follows $\fcnPrio(i_1) < \fcnPrio(i_n)$ which contradicts that there exists an edge $\edgeDirected{i_n}{i_1}$ in the cycle, i.e., $\fcnPrio(i_1) > \fcnPrio(i_n)$.
\end{proof}

\subsection{Graph Coloring}
\label{sec:prso:coloring}
The goal of vertex coloring is to partition a set of vertices $\setVertices$ of a graph into a set of colors $\setColors \subset \mathbb{N}_{>0}$ such that no two adjacent vertices are of the same color.
The mapping of vertices to colors is defined by the function $\varphi \colon \setVertices \to \setColors$.
In order to produce a valid coloring, $\varphi$ has to satisfy
\begin{equation}
    \varphi(i)\neq \varphi(j), \quad \forall i, j \in \setVertices, i \neq j, \edgeDirected{i}{j}\in\setEdges.
\end{equation}
We translate a graph coloring function $\varphi$ to a prioritization function $\fcnPrio_{\text{color}}$. 
Let $\setVertices_c$ be all vertices of color $c\in\setColors$
\begin{equation}
    \setVertices_c = \set{i \mid i \in \setVertices, \varphi(i) = c}.
\end{equation}
We can generate a directed graph from a graph colored with $\varphi$ with a prioritization function $\fcnPrio_{\text{color}}$ that fulfills the requirement
\begin{equation}
    \label{eq:prio_coloring}
    \begin{aligned}
             &&\fcnPrio_{\text{color}}(i) &< \fcnPrio_{\text{color}}(j), \quad \forall i \in \setVertices_{c_1}, \, \forall j \in \setVertices_{c_2} \\
        \iff &&c_1&<c_2.
    \end{aligned}
\end{equation}
By definition of graph coloring, $\fcnPrio_{\text{color}}$ is a valid prioritization function regarding \cref{eq:valid_priority}.
According to \cref{lemma:priority_dag}, the resulting directed graph is a \ac{dag}.

\begin{theorem}\label{th:color_level}
    Let a graph $\graphUndirected$ be colored by the coloring function $\varphi \colon \setVertices \to \setColors$.
    Let $\fcnPrio_{\text{color}}$ be a prioritization function that fulfills \eqref{eq:prio_coloring}, so we can convert a colored coupling graph $\graphUndirected$ to a coupling \ac{dag} $\graphDirected$.
    The number of colors $\numColors$ corresponds to the number of levels $\numLevels$.
\end{theorem}
\begin{proof}
    We proof this by induction. 

    \textbf{Basis:}
    Assume a coupling graph colored with one color.
    This implies that there exist no edges between the vertices of the graph.
    With any $\fcnPrio_{\text{color}}$, it follows that the number of levels is one.
    It can be concluded that the number of colors corresponds to the number of levels for $\numColors=\numLevels=1$ color.
    
    \textbf{Inductive step:}
    We assume that the coupling graph colored with $\numColors=k$ colors results in $k$ levels.
    If a new color is necessary for a vertex, this vertex must be connected with an edge to a vertex of each already assigned color.
    Since there cannot be an edge between vertices in the same level, a new level must be added.
    Therefore, the number of colors is $\numColors=k+1$ and the number of levels is $\numLevels=k+1$.
    It can be concluded that the number of colors corresponds to the number of levels for $k+1$ colors.
\end{proof}

The chromatic number $\chi(\graphUndirected)$ is the smallest number of colors needed to properly color a graph $\graphUndirected$, $\chi \colon \graphUndirected \to \mathbb{N}$.
\begin{problem}\label{prob:coloring}
    Given an undirected coupling graph $\graphUndirected$, find a coloring function $\varphi \colon \setVertices \to \setColors$ that yields a coloring with the chromatic number $\chi(\graphUndirected)$.
\end{problem}
\begin{theorem}
    \Cref{prob:min_level} is equivalent to \cref{prob:coloring} with a prioritization function that respects \eqref{eq:prio_coloring}.
\end{theorem}
\begin{proof}
    Follows directly from \cref{th:color_level}. Since the number of colors in the coupling graph corresponds to the number of levels in the coupling \ac{dag}, minimizing the number of colors is the same as minimizing the number of levels.
\end{proof}

The graph coloring problem belongs to the class of NP-complete problems.
Hence, no efficient algorithm that results in a coloring function for which it holds that $\numColors = \chi(\graphUndirected)$ is known \citep{garey1976complexity}.
Further, the problem of determining the chromatic number of a graph $\chi(\graphUndirected)$ is NP-hard \citep{mcdiarmid1979determining}.
Only special cases exist where the chromatic number is known beforehand, e.g., a 2-coloring for trees or an $\numAgents$-coloring for fully connected graphs.
Hence, we propose a polynomial-time decentralized greedy graph coloring algorithm.

\subsection{Polynomial-time Decentralized Graph Coloring}
\label{sec:prso:algo}
In this section we describe our approach of graph coloring.
In contrast to approaches from literature \citep{kuwata2007distributed, welsh1967upper}, we propose a decentralized algorithm.
The input to the algorithm is the undirected coupling graph, in which every vertex is associated with a unique number.
Every agent in the \ac{ncs} solving this algorithm must obtain the same prioritization to realize decentral execution.

To overcome the efficiency problem of optimal graph coloring algorithms, we approximate the solution to the minimal graph coloring with a greedy algorithm such that $\numColors/\chi(\graphUndirected)$ is close to $1$ \citep{garey1976complexity}.
Greedy coloring algorithms select and color vertices one after another with the corresponding smallest possible color $c \in \setColors$.

To select the next vertex to be colored, we propose a heuristic based on a combination of \iac{sdo}, \iac{ldo}, and \iac{ffo} \citep{al-omari2006new,hasenplaugh2014ordering} to achieve a consistent and near-optimal coloring among agents.
It arranges vertices corresponding to \iac{sdo}, \iac{ldo}, and \iac{ffo}, in descending importance for the ordering.
\Iac{sdo} arranges the vertices in descending order according to the saturation degree, i.e., the number of their differently colored neighbors.
\Iac{ldo} arranges the vertices in descending order according to the number of neighbors $\vertexDegree{i}$.
\Iac{ffo} arranges the vertices by their unique number.
For decentralized execution, the result of our algorithm must be consistent across all agents.
The first two orderings can be ambiguous, if multiple vertices have the same saturation degree and the same number of neighbors.
However, such ambiguities are always resolved through unique vertex numbers, guaranteeing a consistent coloring among the agents.
Consequently, the algorithm results in the same coloring for all agents, given the same input.

\cref{alg:heuristic_greedy_coloring} details our coloring algorithm.
The algorithm iteratively colors vertices until every vertex is colored (\cref{alg:hgc:while}).
All vertices yet to be colored are investigated in the for-loop in \cref{alg:hgc:for}.
If the current saturation degree $s$ is greater than the current maximum saturation degree $s_\text{max}$, $s_\text{max}$ and the next vertex to be colored $i_\text{max}$ are updated (\cref{alg:hgc:update-s,alg:hgc:update-i}).
This procedure corresponds to \iac{sdo}.
If the saturation degrees are equal, the algorithm returns to \iac{ldo} (\cref{alg:hgc:ldo}) and updates $i_\text{max}$ with $i$ if $i$ has a higher degree than the current $i_\text{max}$.
Otherwise, $i_\text{max}$ remains the first vertex in the list of vertices according to \iac{ffo}.
The set of valid colors for vertex $i_\text{max}$ is the set difference of all colors and its adjacent colors (\cref{alg:hgc:adj,alg:hgc:poss}).
The minimal possible color is assigned to $i_\text{max}$ with $\varphi$ (\cref{alg:hgc:assign}), and the vertex is removed from the set of uncolored vertices (\cref{alg:hgc:rem}).
Since in each iteration of the outer loop one vertex is colored, the algorithm will terminate in $\numAgents$ iterations.
Since the number of available colors is infinite, the algorithm will always return a valid coloring.

\begin{algorithm}
    \caption{Decentralized greedy graph coloring}
    \label{alg:heuristic_greedy_coloring}
    \begin{algorithmic}[1]
        \Require Graph $\graphUndirected = (\setVertices, \setEdges)$
        \Ensure Graph coloring $\varphi$
        \State $\setColors \gets \set{1, \ldots, \numAgents}$ \Comment{set of colors}
        \State $\varphi(\setVertices) \gets 0$
        \While{$\setVertices \neq \emptyset$} \label{alg:hgc:while}
            \State $s_\text{max} \gets -1$ \Comment{implicit selection via \ac{ffo} follows}
            \For{$i \in \setVertices$} \label{alg:hgc:for}
                \State $s \gets \lvert \set{c \mid \varphi(j) = c \neq 0, j \in \setNeighbors{i}} \rvert$ \label{alg:hgc:sdo}
                \newline \Comment{saturation degree}
                \If{$s > s_\text{max}$} \label{alg:hgc:gr}
                    \State $s_\text{max} \gets s$ \label{alg:hgc:update-s}
                    \State $i_\text{max} \gets i$ \Comment{selection via \ac{sdo}} \label{alg:hgc:update-i}
                \EndIf
                \If{$s = s_\text{max}$ {\bfseries and} $\vertexDegree{i} > \vertexDegree{i_\text{max}}$} \label{alg:hgc:ldo}
                    \State $i_\text{max} \gets i$  \Comment{selection via \ac{ldo}}
                \EndIf
            \EndFor
            \State $\setColors_\text{adj} \gets \set{c \mid \varphi(j) = c \neq 0, j \in \setNeighbors{i_\text{max}} }$ \label{alg:hgc:adj} \newline \Comment{adjacent colors}
            \State $\setColors_\text{poss} \gets \setColors \setminus \setColors_\text{adj}$ \label{alg:hgc:poss}
            \State $\varphi(i_\text{max}) \gets \min \setColors_\text{poss}$ \label{alg:hgc:assign}
            \State $\setVertices \gets \setVertices \setminus i_\text{max}$ \label{alg:hgc:rem}
        \EndWhile
        \State \Return $\varphi$
    \end{algorithmic}
\end{algorithm}

\Iac{ffo} needs to iterate only once over the set of vertices, resulting in a time complexity of $\mathcal{O}(\numAgents)$. 
\Iac{ldo} needs to compare the degree of every vertex, leading to a time complexity of $\mathcal{O}(\numAgents^2)$.
For every vertex, \iac{sdo} needs to check every adjacent vertex to determine the saturation degree, which results in a time complexity of $\mathcal{O}(\numAgents^3)$.
Thus, \cref{alg:heuristic_greedy_coloring} also has a polynomial time complexity of $\mathcal{O}(\numAgents^3)$.

\begin{remark}
    Assuming edges in the coupling graph contain constraints, their number equals the number of incoming edges $\vertexInDegree{i}$ in \ac{pp}, which depends on the prioritization function $\fcnPrio$.
    We can reorder the sequence of colors and thus the computation levels to influence the number of incoming edges of vertices.
    Equally distributing the number of constraints among agents can be beneficial for the feasibility of the planning problems and the solution cost of the \ac{mamp} instance.
    A vertex with a high degree should therefore be on a level with a low number, so that the edges are outgoing rather than incoming.
\end{remark}

\begin{figure}
    \centering
    \begin{subfigure}{\linewidth}
        \centering
            \begin{tikzpicture}[roundnode/.style={circle, draw, minimum size=7mm}, baseline=(current bounding box.center), node distance=1.5cm]
        \node[roundnode] (1) {$1$};
        \node[roundnode] (2) [below right of=1] {$2$};
        \node[roundnode] (3) [below left of=2] {$3$};
        \node[roundnode] (4) [below left of=1] {$4$};

        \draw[] (1) -- (2);
        \draw[] (2) -- (3);
        \draw[] (3) -- (4);
        \draw[] (4) -- (1);
        \draw[] (4) -- (2);
    \end{tikzpicture}
        \subcaption{Undirected coupling graph}
        \label{subfig:process_1_undirected}
    \end{subfigure}
    \par\bigskip
    \begin{subfigure}{\linewidth}
        \centering
            \begin{tikzpicture}[
        roundnode/.style={circle, draw, minimum size=7mm, preaction={fill=white}},
        baseline=(current bounding box.center),
        node distance=1.5cm
        ]

        \node[roundnode, fill = white] (1) {\contour{white}{$1$}};
        \node[roundnode, fill = white] (2) [below = 1 of 1] {\contour{white}{$2$}};
        \node[roundnode, fill = white] (3) [below = 1 of 2] {\contour{white}{$3$}};
        \node[roundnode, fill = white] (4) [below = 1 of 3] {\contour{white}{$4$}};

        \draw[-latex] (1) to (2);
        \draw[-latex, bend left] (1) to (4);
        \draw[-latex] (2) to (3);
        \draw[-latex, bend right] (2) to (4);
        \draw[-latex] (3) to (4);

        \node[roundnode, pattern color = color5, pattern = {horizontal lines}]     (c1) [right=1.5 of 1]   {\contour{white}{$1$}};
        \node[roundnode, pattern color = color5, pattern = {horizontal lines}]     (c3) [right=2.5 of 1]   {\contour{white}{$3$}};
        \node[roundnode, pattern color = color1, pattern = {vertical lines}]   (c2) [right=2 of 2]     {\contour{white}{$2$}};
        \node[roundnode, pattern color = color2, pattern = {north east lines}]    (c4) [right=2 of 3]     {\contour{white}{$4$}};
        
        \draw[-latex] (c1) to (c2);
        \draw[-latex, bend right] (c1) to (c4);
        \draw[-latex] (c3) to (c2);
        \draw[-latex, bend left] (c3) to (c4);
        \draw[-latex] (c2) to (c4);

        \node[roundnode, pattern color = color1, pattern = {vertical lines}]   (ro2) [right=4.5 of 1]     {\contour{white}{$2$}};
        \node[roundnode, pattern color = color2, pattern = {north east lines}]    (ro4) [right=4.5 of 2]     {\contour{white}{$4$}};
        \node[roundnode, pattern color = color5, pattern = {horizontal lines}]     (ro1) [right=4 of 3]   {\contour{white}{$1$}};
        \node[roundnode, pattern color = color5, pattern = {horizontal lines}]     (ro3) [right=5 of 3]   {\contour{white}{$3$}};
        
        \draw[-latex] (ro2) to (ro4);
        \draw[-latex, bend right] (ro2) to (ro1);
        \draw[-latex, bend left] (ro2) to (ro3);
        \draw[-latex] (ro4) to (ro1);
        \draw[-latex] (ro4) to (ro3);

        \node (l1) [left=1 of 1] {CL 1};
        \node (l2) [left=1 of 2] {CL 2};
        \node (l3) [left=1 of 3] {CL 3};
        \node (l4) [left=1 of 4] {CL 4};

        \node (r1) [right=5.7 of 1] {};
        \node (r2) [right=5.7 of 2] {};
        \node (r3) [right=5.7 of 3] {};
        \node (r4) [right=5.7 of 4] {};

        \begin{pgfonlayer}{bg}
            \draw[loosely dashed] (l1) -- (r1);
            \draw[loosely dashed] (l2) -- (r2);
            \draw[loosely dashed] (l3) -- (r3);
            \draw[loosely dashed] (l4) -- (r4);
        \end{pgfonlayer}
    \end{tikzpicture}
        \subcaption{Computation levels (CL) and coupling \ac{dag}; left: priorities equal to vertex number; middle: priorities based on graph coloring; right: reordered colors for reduced maximum number of incoming edges per computation level (our approach).}
        \label{subfig:process_2_directed}
    \end{subfigure}
    \caption{Example of problem solution process}
    \label{fig:process}
\end{figure}
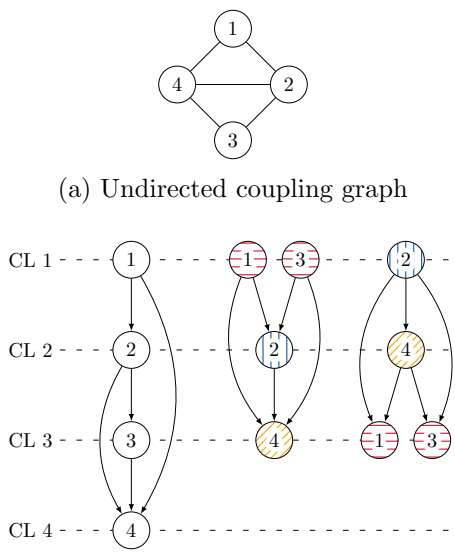

\Cref{fig:process} illustrates the proposed problem solution with an example.
From an example undirected graph (\cref{subfig:process_1_undirected}), a baseline prioritization which assigns priorities equal to the vertex number results in four computation levels.
Coloring the vertices solves \cref{prob:min_level} and reduces the number of computation levels to three.
Further reordering of the levels reduces the maximum number of incoming edges per agent from two to one and from three to two.

\newcommand{\evalTimeReduction}{57.9}%
\newcommand{\evalCostIncrease}{26.2}%

\section{Evaluation in a Trajectory Planning Application of \acp{cav}}
\label{sec:evaluation}
We evaluate the presented approach on reducing computation time by prioritization in the context of \acp{cav}.
In \cref{sec:eval:intersection}, we present the experiment setup, an intersection scenario with eight vehicles.
In \cref{sec:eval:computation_time}, we compare the computation time of the \ac{mamp} instance of our approach with prioritization approaches from literature.
In \cref{sec:eval:quality}, we evaluate the quality of the trajectories of the vehicles in the \ac{mamp} instance.

The algorithms are implemented in MATLAB and are publicly available%
\footnote{\href{https://github.com/embedded-software-laboratory/p-dmpc}{github.com/embedded-software-laboratory/p-dmpc}}%
.
They ran on an off-the-shelf laptop with an Apple Silicon M3 Pro 12-Core CPU with \qtyrange{2.7}{4.1}{\giga\hertz} and \qty{36}{\giga\byte} of RAM. 

We compare our prioritization algorithm with three algorithms from literature.
Each algorithm is represented by a prioritization function $\fcnPrio \colon \setVertices \to \mathbb{N}$.
The first function prioritizes vehicles according to their vertex number \citep{alrifaee2016coordinated} and is denoted by $\fcnPrio_\text{constant}$.
The second function prioritizes vehicles randomly at each time step \citep{bennewitz2002finding} and is denoted by $\fcnPrio_\text{random}$.
The third function prioritizes vehicles with a constraint-based heuristic \citep{scheffe2022increasing}.
A higher number of potential collisions with other vehicles result in a higher priority.
The function is denoted by $\fcnPrio_\text{constraint}$.

\subsection{Trajectory Planning}
\label{sec:eval:trajectory-planning}
In our application of trajectory planning for \acp{cav}, the objective \cref{eq:pdmpc:ocp_obj} of a \ac{cav} $\anAgent$ is to stay close to a reference trajectory.
In our work, the vector field $f_d^{(i)}$ is a nonlinear kinematic single-track model \citep[section~2.2]{rajamani2006vehicle}.
The coupling constraint \cref{eq:pdmpc:ocp_c_coupling_sequential} achieves collision avoidance with predecessors.

It is computationally hard to find the global optimum to \ac{ocp} \cref{eq:4-ocp} due to its nonlinearity and nonconvexity.
We approximate \cref{eq:4-ocp} with a receding horizon graph search based on our previous work \citep{scheffe2023receding} which can be solved online.
The system model is quantized in \iac{mpa} which consists of a set of states and a set of transitions, called \acp{mp}.
Finding a feasible solution to \cref{eq:4-ocp} results in a tree search which can be solved by means of an A$^*$ algorithm.
In this work, we use a sampling-based approach on the basis of \ac{mcts}.
In each time step, our \ac{mcts} evaluates $N_{\text{exp}}$ random transitions in the search tree.
The algorithm thus builds part of the complete search tree.
After $N_{\text{exp}} \in \setNaturalNumbers$ transitions, the path with the lowest cost according to \cref{eq:pdmpc:ocp_obj} is selected as the solution to \cref{eq:4-ocp}.
The algorithm has no guarantee on optimality, but achieves feasible solutions with nearly constant computation time, satisfying \cref{ass:equal_comp_time}.

\subsection{Intersection Scenario}
\label{sec:eval:intersection}
\begin{figure}
    \centering
    \includegraphics{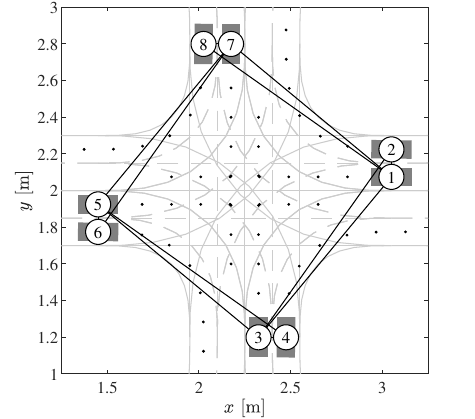}
    \caption{
        Coupling graph for eight-lane intersection with eight vehicles.
        For each direction of the intersection, one vehicle turns right and one vehicle moves straight, which is indicated by dots.
        The undirected coupling graph is based on the potential collisions between vehicles.
    }
    \label{fig:eval_scenario_coupling}
\end{figure}

We evaluate the presented approach to reduce computation time by prioritization to distributed trajectory planning for road vehicles at an intersection with two incoming and two outgoing lanes for each direction.
The intersection is part of a simulation of the \ac{cpmlab}, an open-source, remotely-accessible testbed for \acp{cav} \citep{kloock2021cyberphysical}.
\cref{fig:eval_scenario_coupling} depicts the initial states with eight vehicles as well as the associated undirected coupling graph.
From each direction, one vehicle is turning right, and one vehicle is moving straight.
The experiments run for $\timestep_{\text{experiment}}=25$ time steps with a time step duration of $\qty{0.2}{s}$.
The \ac{mpc} uses a prediction horizon of $\horizonPrediction=8$, the \ac{mcts} which solves the \ac{ocp} uses $N_\text{exp}=\num{2500}$ expansions.

\subsection{Computation Time}
\label{sec:eval:computation_time}
\begin{figure}
    \centering
    \includegraphics{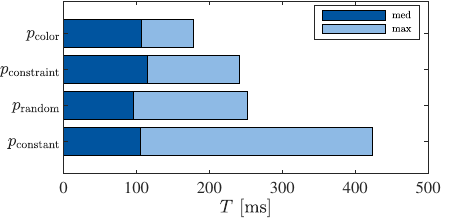}
    \caption{
        Computation time using $\fcnPrio_\text{color}$ (our approach) compared to other prioritizations in the experiment starting from the initial states shown in \cref{fig:eval_scenario_coupling}.
    }
    \label{fig:eval_time}
\end{figure}

\Cref{fig:eval_time} shows the computation time of the \ac{mamp} instance according to \cref{eq:computation_time} for different prioritizations.
For each prioritization, the median and maximum computation time over all time steps are displayed.
In our experiment, we decrease the maximum computation time by more than \qty{\evalTimeReduction}{\percent} compared to the prioritization $\fcnPrio_\text{constant}$.

\begin{figure}
    \centering
    \includegraphics{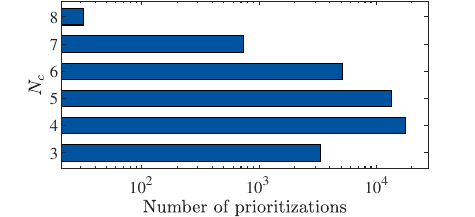}
    \caption{
        Number of prioritizations for the coupling graph in \cref{fig:eval_scenario_coupling} that result in a coupling \ac{dag} with the given number of computation levels $\numLevels$.
    }
    \label{fig:eval_level_distribution}
\end{figure}

The decrease in computation time comes mostly from a reduction in the number of computation levels.
Each prioritization produces a directed coupling graph, which results in a certain number of computation levels.
As the number of different prioritizations for $\numAgents$ vehicles is $\numAgents !$, there are $8! = \num{40320}$ different prioritizations for the vehicles.\newline%
\cref{fig:eval_level_distribution} shows the number of computation levels for these prioritizations, which ranges from three to eight given the undirected coupling graph in \cref{fig:eval_scenario_coupling}.

\begin{figure}
    \centering
    \includegraphics{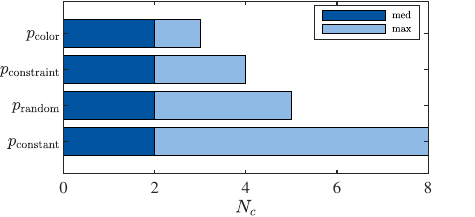}
    \caption{
        Number of computation levels using $\fcnPrio_\text{color}$ (our approach) compared to other prioritizations in the experiment starting from the initial states shown in \cref{fig:eval_scenario_coupling}.
    }
    \label{fig:eval_levels}
\end{figure}

\Cref{fig:eval_levels} shows the median and maximum computation levels $\numLevels$ over all time steps.
As vehicles move during the experiment, the undirected coupling graph changes, and with it the number of computation levels.
The median is at two, since as vehicles exit the intersection, only two vehicles are coupled in each outgoing lane.
With our algorithm to solve the \ac{ocp} (\cref{sec:eval:trajectory-planning}), we obtain a strong correlation between the number of computation levels and the computation time of the \ac{mamp} instance.
The maximum number of computation levels for $p_\text{random}$ is five, which can be expected given most prioritizations result in four or five computation levels (\cref{fig:eval_level_distribution}).

\begin{figure}
    \centering
    \includegraphics{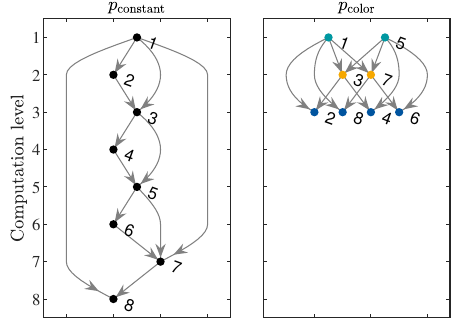}
    \caption{
        Computation levels and coupling \ac{dag} with prioritizations $p_\text{constant}$ and $p_\text{color}$ (our approach) for the coupling graph in \cref{fig:eval_scenario_coupling}
    }
    \label{fig:eval_dag}
\end{figure}

\cref{fig:eval_dag} depicts the directed coupling graphs resulting from $p_\text{constant}$ and our proposed coloring prioritization function $\fcnPrio_\text{color}$ for the first time step of the experiment with the undirected coupling graph depicted in \cref{fig:eval_scenario_coupling}.
Prioritization with $p_\text{constant}$ results in eight computation levels, whereas we can color the graph using \cref{alg:heuristic_greedy_coloring} with only three different colors, resulting in three computation levels.

The vehicles compute the prioritization with our approach in a decentralized fashion and thus parallelly.
The maximum of the graph coloring computation time is \SI{0.81}{\milli\second}, which is negligible when compared to the trajectory planning task.

\subsection{Quality of Trajectories}
\label{sec:eval:quality}
\begin{figure}
    \centering
    \includegraphics{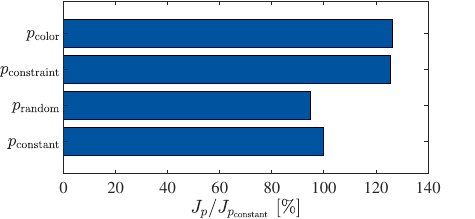}
    \caption{
        Cost $\fcnObjective_{\fcnPrio_\text{color}}$ using our approach $\fcnPrio_\text{color}$ compared to the cost $\fcnObjective_{\fcnPrio}$ using other prioritizations in the experiment starting from the scene shown in \cref{fig:eval_scenario_coupling}.
        Cost is normalized to the cost $\fcnObjective_{\fcnPrio_\text{constant}}$ of prioritization $\fcnPrio_\text{constant}$.
    }
    \label{fig:eval_cost}
\end{figure}

The quality of trajectories $\fcnObjective_{\fcnPrio}$ of a prioritization algorithm $\fcnPrio$ according to the cost of the \ac{ocp} of each vehicle is given as
\begin{equation}
        \fcnObjective_{\fcnPrio}
        = \sum_{\timestep=0}^{\timestep_{\text{experiment}}} \sum_{\anAgent=1}^{\numAgents} \fcnObjective[\anAgent]_{\fcnPrio}(\timestep),
\end{equation}
with $\timestep_{\text{experiment}}$ being the number of time steps in the experiment, and $\fcnObjective[\anAgent]_{\fcnPrio}(\timestep)$ given in \cref{eq:objective}.
The cost normalized to that of the prioritization $p_\text{constant}$ is shown in \cref{fig:eval_cost}.
In our experiment, the cost increased by \qty{\evalCostIncrease}{\percent} compared to the prioritization $p_\text{constant}$.
All vehicles were able to pass the intersection with all prioritizations.

\section{Discussion}
\label{sec:discussion}
In the following, we discuss our approach which prioritized agents using graph coloring.
The discussion regards the computation effort, the quality of agents' solutions in the \ac{mamp} instance, and the communication effort.

\subsection{Computation Effort}
The computation time of our coloring algorithm scales in an order of $\mathcal{O}(\numAgents^3)$ with the number of agents as discussed in \cref{sec:prso:coloring}.
This effort should at least be compensated by the reduced computation time for the \ac{mamp} instance.
The amount of reduced computation time for the \ac{mamp} instance depends on mainly two factors.

The first factor is the computation demand for the control problem in each agent.
If the control problem in each agent is computationally simple, the reduction of computation levels might not be very beneficial.
However, nonconvex optimization problems, such as trajectory planning with collision avoidance, typically are computationally demanding and therefore benefit from a reduction of computation levels.
If the computation demand is sensitive to the prioritization and thus the constraints, which can be the case in A$^*$-based algorithms, the reduction in computation levels might be counteracted by an increase in computation demand.
In sampling-based algorithms like \ac{mcts}, the nearly constant computation demand mainly depends on the number of explored samples.
Thus, the reduction of computation time is proportional to the reduction of computation levels.

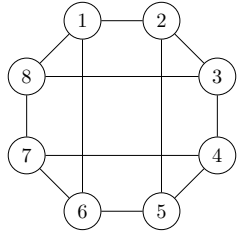
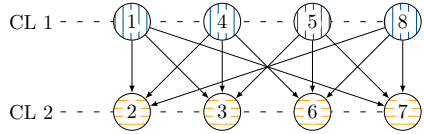
\begin{figure}
    \centering
    \begin{subfigure}{\linewidth}
        \centering
            \begin{tikzpicture}[roundnode/.style={circle, draw, minimum size=7mm}, baseline=(current bounding box.center), node distance=1.5cm]
        \node[roundnode] (1) {$1$};
        \node[roundnode] (2) [right of=1] {$2$};
        \node[roundnode] (3) [below right of=2] {$3$};
        \node[roundnode] (4) [below of=3] {$4$};
        \node[roundnode] (5) [below left of=4] {$5$};
        \node[roundnode] (6) [left of=5] {$6$};
        \node[roundnode] (7) [above left of=6] {$7$};
        \node[roundnode] (8) [above of=7] {$8$};

        \draw (1) -- (2);
        \draw (2) -- (3);
        \draw (3) -- (4);
        \draw (4) -- (5);
        \draw (5) -- (6);
        \draw (6) -- (7);
        \draw (7) -- (8);
        \draw (8) -- (1);
        
        \draw (1) -- (6);
        \draw (2) -- (5);

        \draw (3) -- (8);
        \draw (4) -- (7);
    \end{tikzpicture}
        \subcaption{Undirected bipartite coupling graph}
        \label{subfig:eval_8theosetup}
    \end{subfigure}
    \par\bigskip
    \begin{subfigure}{\linewidth}
        \centering
            \begin{tikzpicture}[
        roundnode/.style={circle, draw, minimum size=7mm, preaction={fill=white}},
        baseline=(current bounding box.center),
        node distance=1.5cm
    ]
        \node[roundnode, pattern color = color1, pattern = {vertical lines}] (c1) {\contour{white}{$1$}};
        \node[roundnode, pattern color = color2, pattern = {horizontal lines}] (c2) [below=1 of c1] {\contour{white}{$2$}};
        \node[roundnode, pattern color = color2, pattern = {horizontal lines}] (c3) [right=1 of c2] {\contour{white}{$3$}};
        \node[roundnode, pattern color = color1, pattern = {vertical lines}] (c4) [right=1 of c1] {\contour{white}{$4$}};
        \node[roundnode, pattern color = color1, pattern = {vertical lines}] (c5) [right=1 of c4] {\contour{white}{$5$}};
        \node[roundnode, pattern color = color2, pattern = {horizontal lines}] (c6) [right=1 of c3] {\contour{white}{$6$}};
        \node[roundnode, pattern color = color2, pattern = {horizontal lines}] (c7) [right=1 of c6] {\contour{white}{$7$}};
        \node[roundnode, pattern color = color1, pattern = {vertical lines}] (c8) [right=1 of c5] {\contour{white}{$8$}};

        \draw[-latex] (c1) -- (c2);
        \draw[-latex] (c1) -- (c3);
        \draw[-latex] (c1) -- (c7);

        \draw[-latex] (c4) -- (c2);
        \draw[-latex] (c8) -- (c2);

        \draw[-latex] (c4) -- (c3);
        \draw[-latex] (c5) -- (c3);

        \draw[-latex] (c4) -- (c6);

        \draw[-latex] (c5) -- (c6);
        \draw[-latex] (c5) -- (c7);

        \draw[-latex] (c8) -- (c6);

        \draw[-latex] (c8) -- (c7);

        \node (l1) [left=1 of c1] {CL 1};
        \node (l2) [left=1 of c2] {CL 2};

        \node (r1) [right=6.5 of l1] {};
        \node (r2) [right=6.5 of l2] {};

        \begin{pgfonlayer}{bg}
            \draw[loosely dashed] (l1) -- (r1);
            \draw[loosely dashed] (l2) -- (r2);
        \end{pgfonlayer}
    \end{tikzpicture}
        \subcaption{Computation levels (CL) and coupling \ac{dag}}
        \label{subfig:eval_8theotopo}
    \end{subfigure}
    \caption{Transformation of undirected bipartite graph to \ac{dag} using graph coloring}
    \label{fig:eval_8theotopo}
\end{figure}

The second factor is the undirected coupling graph of the \ac{mamp} instance.
For a fully connected coupling graph, any prioritization algorithm results in $\numAgents$ computation levels.
For a coupling graph which is a path graph, the worst prioritization with regards to the computation time would result in $\numAgents$ levels, whereas our prioritization through graph coloring always results in only two levels.
A path graph is a graph that consists of a single path \citep{lunze2019networked} in which each vertex has a degree of two except the vertices at the end and beginning of the path have a degree of one.
Coupling graphs that have a chromatic number of two, such as the one shown in \cref{fig:eval_8theotopo}, are bipartite graphs.
For the graph in \cref{fig:eval_8theotopo}, a prioritization with priorities according to vertex numbers would result in eight computation levels, while our proposed coloring algorithm achieves two computation levels.
\begin{theorem}\label{thm:minlevels-maxdegree}
    Given an undirected coupling graph $\graphUndirected$, the minimum number of computation levels $\numLevels$ is upper bounded by the maximum degree of vertices in the graph, i.e.,
    \begin{equation}
        \min_{\fcnPrio} \numLevels (\graphUndirected, \fcnPrio) \leq \max_{\anAgent\in\setAgents} \vertexDegree{\anAgent} + 1.
    \end{equation}
\end{theorem}
\begin{proof}
    The number of computation levels $\numLevels$ corresponds to the longest path in the directed coupling graph $\graphDirected$. By the \cite{gallai1968directed}-\cite{hasse1965zur}-\cite{roy1967nombre}-\cite{vitaver1962determination} theorem,
    \begin{equation}
        \min_{\fcnPrio} \numLevels (\graphUndirected, \fcnPrio) = \chi(\graphUndirected).
    \end{equation}
    According to \cite{welsh1967upper},
    \begin{equation}
        \chi(\graphUndirected) \leq \max_{\anAgent\in\setAgents} \vertexDegree{\anAgent} + 1.
    \end{equation}
\end{proof}
From \cref{thm:minlevels-maxdegree} follows that the number of computation levels, and thus the networked computation time, is not depending on the number of agents in the coupling graph, but only on its maximum degree.
From \cref{thm:minlevels-maxdegree} we also conclude that sparse coupling graphs with few edges are more likely to benefit from our approach.
The maximum number of computation levels that can result from any prioritization is equal to the length of the longest path in the undirected coupling graph.
The greater the difference of the chromatic number and the length of the longest path in the undirected coupling graph, the higher the potential computation time reduction for the \ac{mamp} instance with a prioritization through graph coloring.

\subsection{Quality of Solutions}
The quality of a solution consists of its cost, and its feasibility, i.e., does a solution exist which satisfies all constraints.
Our approach does not need any application-specific knowledge about agents' constraints or interactions.
On the one hand, this makes the approach more general.
On the other hand, the approach cannot consider constraints or interactions during prioritization, which might result in solutions of lower quality.
Our approach prioritizes agents at each time step, which can result in an inversion of the priorities between agents.
Such inversions can drastically change the \ac{ocp} of an agent.
This stands in disharmony to the concept of \ac{mpc}, on which our \ac{pp} approach for \ac{mamp} relies on.
In \ac{mpc} the \ac{ocp} is expected to be known for the prediction horizon.
Consequently, frequent changes of priorities might deteriorate the solution quality.
Additionally, our cost metric takes the whole optimized solution of agents into account, even though only the first control input is applied.
In our experiment, we observed an increase of solution cost by \qty{\evalCostIncrease}{\percent} compared to the prioritization $p_\text{constant}$.
However, the random prioritization $p_\text{random}$, which also frequently changes priorities, increases the solution quality compared to $p_\text{constant}$.
Thus, in general, it seems difficult to assess the solution quality on the basis of the coupling \ac{dag} structure.
With our proposed approach, all vehicles were able to cross the intersection.
The solution quality should be seen in relation to the highly decreased computation time.

\subsection{Communication Effort}
The communication in our \ac{pp} framework (\cref{fig:pp_framework}) is limited to communicating predictions to successors, i.e., communication takes place where agents are connected in the directed coupling graph.
The number of sequential communications corresponds to the number of computation levels.
Therefore, through reducing the number of computation levels, our approach also reduces the number of sequential communications.
However, communication instead takes place in parallel, which puts a higher communication load on the communication infrastructure.
Depending on the infrastructure, the implications on the communication can be beneficial or detrimental.

\section{Conclusion}
\label{sec:conclusion}
We proposed a prioritization strategy which reduces the computation time in \ac{pp} for \ac{mamp}, a variant of \ac{mapf}.
We proved the equivalence of this problem to graph coloring of a given undirected coupling graph of the \ac{mamp} instance and prioritizing agents on the basis of the vertex colors.
Our coloring algorithm works in a decentralized fashion, which avoids a single point of failure and reduces the amount of communication required.
We successfully applied our process for prioritization to vehicles crossing an eight-lane intersection using an \ac{pp} approach for \ac{mamp} based on \ac{mpc}, for which the reduction of computation time is around \qty{\evalTimeReduction}{\percent} compared to a baseline approach from literature.
Our approach can be applied to any domain with prioritized computations, provided that the coupling graph is known.
Especially in large-scale systems with sparse coupling graphs, our approach can significantly reduce the computation time and thus improve the scalability.

The feasibility of each agent's planning problem and the quality of the solutions also depend on the prioritization.
In our experiment, the solution cost was increased by \qty{\evalCostIncrease}{\percent} compared to a baseline approach from literature.
In the future, we will further investigate prioritization considering feasibility and quality of solutions.

\section{Acknowledgements}
This research was funded by the Deutsche Forschungsgemeinschaft (DFG, German Research Foundation) within the Priority Program SPP 1835 ``Cooperative Interacting Automobiles'' (grant number: KO 1430/17-1).

\vskip 0.2in
\bibliography{submodules/symbols/GROKO,submodules/symbols/scheffe-publications}
\bibliographystyle{apacite}

\end{document}